\documentclass[12pt]{article}

\usepackage{amssymb}
\usepackage{amsmath}
\usepackage{graphics,graphicx}

\newcommand{\beq}{\begin{equation}}
\newcommand{\eeq}{\end{equation}}
\newcommand{\lb}{\label}
\newcommand{\beqar}{\begin{eqnarray}}
\newcommand{\eeqar}{\end{eqnarray}}
\newcommand{\bit}{\begin{itemize}}
\newcommand{\eit}{\end{itemize}}
\newcommand{\barr}{\begin{array}}
\newcommand{\earr}{\end{array}}
\def\ds{\displaystyle}

\def\scalp{\mbox{\boldmath $\, \cdot \,$}}
\def\bob{{\, \underline{\overline{\otimes}} \,}}
\def\gdp{\ \begin{picture}(6,6)\put(0,0){\framebox(6,6){$\times$}}\end{picture} \ }

\newcommand{\deriv}[2]{\frac{\partial #1}{\partial #2}}
\newcommand{\nderiv}[3]{\frac{\partial^{\,#3} #1}{\partial #2^{\,#3}}}

\newcommand{\oldr}[1]{\stackrel{\circ}{\mbox{#1}}}
\newcommand{\pv}{-\mskip-19mu\int}

\def\b0{\mbox{\boldmath $0$}}

\def\bb{\mbox{\boldmath $b$}}

\def\bg{\mbox{\boldmath $g$}}

\def\bk{\mbox{\boldmath $k$}}

\def\bn{\mbox{\boldmath $n$}}

\def\bp{\mbox{\boldmath $p$}}
\def\bq{\mbox{\boldmath $q$}}

\def\bs{\mbox{\boldmath $s$}}

\def\bu{\mbox{\boldmath $u$}}
\def\bv{\mbox{\boldmath $v$}}
\def\bw{\mbox{\boldmath $w$}}
\def\bx{\mbox{\boldmath $x$}}

\def\bA{\mbox{\boldmath $A$}}
\def\bB{\mbox{\boldmath $B$}}
\def\bC{\mbox{\boldmath $C$}}
\def\bD{\mbox{\boldmath $D$}}

\def\bF{\mbox{\boldmath $F$}}
\def\bG{\mbox{\boldmath $G$}}

\def\bK{\mbox{\boldmath $K$}}
\def\bL{\mbox{\boldmath $L$}}

\def\bP{\mbox{\boldmath $P$}}
\def\bQ{\mbox{\boldmath $Q$}}

\def\bS{\mbox{\boldmath $S$}}
\def\bT{\mbox{\boldmath $T$}}

\def\Id{\mbox{\boldmath $I$}}

\def\f0{\mbox{$\mathbb{O}$}}

\def\fC{\mbox{$\mathbb{C}$}}

\def\fE{\mbox{$\mathbb{E}$}}

\newcommand{\mK}{\ensuremath{\mathcal{K}}}

\newcommand{\mR}{\ensuremath{\mathcal{R}}}

\def\tr{\mbox{$\mathrm{tr}$}}

\def\dev{\mbox{$\mathrm{dev}$}}
\def\div{\mbox{$\mathrm{div}$}}

\def\cos{\mbox{$\mathrm{cos}$}}
\def\sin{\mbox{$\mathrm{sin}$}}

\def\Ci{\mbox{$\mathrm{Ci}$}}
\def\Si{\mbox{$\mathrm{Si}$}}
\def\sgn{\mbox{$\mathrm{sgn}$}}
\def\log{\mbox{$\mathrm{log}$}}

\def\Orth{\mbox{$\mathsf{Orth}$}}
\def\Orth+{\mbox{$\mathsf{Orth^+}$}}

\newcommand{\Reals}{\ensuremath{\mathbf{R}}}

\def\EJMA{{\it Eur.\ J.\ Mech. A-Solids.}\ }

\def\IJNME{{\it Int.\ J.\ Numer.\ Meth.\ Eng.}\ }

\def\IJSS{{\it Int.\ J.\ Solids Struct.}\ }

\def\JMPS{{\it J.\ Mech.\ Phys.\ Solids}\ }

\begin{document}

\title{A dynamical interpretation of flutter instability \\ in a continuous medium}

\author{Andrea Piccolroaz$^{(*)}$, Davide Bigoni$^{(*)}$ and John R. Willis$^{(\circ)}$\\ \\
$^{(*)}$ Dipartimento di Ingegneria Meccanica e \\ Strutturale,
Universit\`a di Trento, \\ Via Mesiano 77, I-38050 Trento, Italia\\
$^{(\circ)}$ Department of Applied Mathematics and Theoretical Physics \\
Centre for Mathematical Sciences, Cambridge University \\
Wilberforce Road, Cambridge CB3 OWA U.K. \\
email: andrea.piccolroaz@ing.unitn.it; bigoni@ing.unitn.it; \\ J.R.Willis@damtp.cam.ac.uk \\
}


\maketitle

\begin{abstract}
\noindent 
Flutter instability in an infinite medium is a form of material instability corresponding to the occurrence 
of complex conjugate squares of the acceleration wave velocities. Although its occurrence is known to be possible 
in elastoplastic materials with nonassociative flow law and to correspond to some dynamically growing 
disturbance, its mechanical meaning has to date still eluded 
a precise interpretation. This is provided here by constructing the infinite-body, time-harmonic 
Green's function for the 
loading branch of an elastoplastic material in flutter conditions. Used as a perturbation, it reveals that 
flutter corresponds to a spatially blowing-up disturbance, exhibiting well-defined directional properties, determined
by the wave directions for which the eigenvalues become complex conjugate.
Flutter is shown to be connected to the formation of localized deformations, a dynamical phenomenon sharing 
geometrical similarities with the 
well-known mechanism of shear banding occurring under quasi-static loading.  
Flutter may occur much earlier than shear banding
in a process of continued plastic deformation.
\end{abstract}


\newpage

\section{Introduction}

Several micromechanisms acting at a microscale during deformation of granular and rock-like materials 
involve Coulomb friction. As a consequence, the flow rule becomes nonassociative and the phenomenological 
rate elastoplastic constitutive equations for these materials become unsymmetric.
Due to this lack of symmetry, two squares of the propagation velocity of acceleration waves or, in other 
words, two eigenvalues of the acoustic tensor, may become a complex conjugate pair.
That this situation might correspond to a form of material instability particularly relevant
in granular material was clear since J.R. Rice (1977) coined for it the term \lq flutter instability', but neither
examples of constitutive equations displaying this instability nor a mechanical interpretation for it were given at 
that time.

Consequently, research was initially focused on the determination of situations in which flutter was possible (see Bigoni, 2000; 
Loret et al., 2000 for reviews).
In particular, it was shown that flutter instability may 
occur more often than one might expect, not satisfying any hierarchical relation to other instabilities 
(such as for instance shear banding or second-order work negativity), possibly at an early stage 
of a hardening process and typically triggered 
by noncoaxiality (of the flow rule or induced by elastic or plastic anisotropy). 
However, the problem of finding a mechanical interpretation for the instability 
remained almost completely unexplored [with the exceptions of Bigoni and Willis (1994) and Sim\~{o}es (1997), the former
considering a very simple problem setting and the latter providing some numerical tests].
This has been a major problem retarding further progress in research since, though generically 
believed to correspond to a dynamically growing disturbance, only 
the knowledge of the precise mechanical features of the instability can permit its identification for real materials.

To shed light on this problem, a perturbative approach is developed in this article, following the methodology
proposed by
Bigoni and Capuani (2002; 2005) to 
investigate shear banding and other forms of material instabilities. 
In more detail, the analysis is limited in the present article to the loading 
branch\footnote{See Bigoni and Petryk (2002) for a discussion of this delicate 
assumption.} of an elastoplastic constitutive operator (taken from Bigoni and Petryk, 2002) embodying features typical of the 
behaviour of granular materials and capable of exhibiting flutter instability. An infinite body 
is considered made up of this material, homogeneously and 
quasi-statically deformed in two dimensions (plane strain or generalized plane stress). For this  
configuration a time-harmonic Green's function is found (in the way shown by Willis, 1991), which represents 
the first dynamic Green's function obtained for a nonsymmetric constitutive equation\footnote{A quasi-static 
Green's function 
for unsymmetric constitutive equation has been developed by Bertoldi et al. (2005), but this is unsuitable 
for flutter 
analyses, since this instability is essentially dynamic 
and thus remains unrevealed under the quasi-static assumption. In addition, Bertoldi et al. (2005) 
also derive 
boundary integral equations under the unsymmetric constitutive assumption, which are shown to possess
certain typical features although not directly connected to the present discussion.}.
The Green's function is employed to form a pulsating dipole 
(two equal and opposite forces 
having a magnitude varying sinusoidally with time) to be used 
as a dynamic perturbation revealing effects of flutter. 

Results demonstrate the following features of flutter instability that may\footnote{More precisely, 
flutter instability 
has been shown by Bigoni and Loret (1999) to be unrelated to the occurrence 
of other instabilities such as loss of positive definiteness
of second order work, loss of strong ellipticity and loss of ellipticity.}
also occur in a material for which 
{\em the tangent constitutive operator is positive definite} 
(so that negative second-order work and shear bands are excluded
at the considered stress level).

\begin{itemize}

\item Differently from shear bands, becoming already evident when the boundary of the region of 
ellipticity is approached from its interior (Bigoni and Capuani, 2002; 2005), flutter 
instability remains undetected while
the eigenvalues of the acoustic tensor lie in the real range, appearing only after two real eigenvalues have
coalesced and then become a complex conjugate pair;

\item flutter instability corresponds to a disturbance blowing-up in space from the perturbing 
dipole and {\em self-organizing along well-defined plane waves}. 

\item the normals to the above plane waves lie within the fan of 
directions corresponding to flutter and have been found to have an inclination remarkably different 
from that corresponding to shear bands, occurring later in the hardening process.

\end{itemize}

It should be noted that 
the blow-up found in our solution will occur rapidly and nonlinearities neglected in our analysis (such as for
instance the possibility of elastic unloading and plastic reloading)
may soon become important, possibly changing the overall mechanical response.
Equally significant is the fact that the rate of growth increases with the frequency that is adopted. The governing
equations of motion thus represent a problem that is {\em dynamically ill-posed} in the general transient case,
{\em unless} the tangent moduli in fact display a frequency-dependence, such that the flutter effect
reduces as frequency increases\footnote{Such a model was introduced by Bigoni and Willis (1994) in the context of a simple
one-dimensional example.}. 
However, our results 
suggest that flutter instability should induce a layering in an initially homogeneous material, inducing 
a localization of strain in a 
form somehow similar ---though possibly occurring much earlier in a hardening process--- 
to that pertaining to shear bands occurring in a dynamical context (Bigoni and Capuani, 2005).
Our hope is that this feature revealed by our results has now been made accessible to experimental investigation.

\subsection{Notation}
A standard, intrinsic notation is used throughout the paper (as for instance in Bigoni and Loret, 1999 and 
Bigoni, 2000), where vectors and second-order tensors are denoted by 
bold (the latter capital) letters. The scalar product between arbitrary tensors $\bA$ and $\bB$ is denoted by 
\beq
\bA\scalp\bB = \tr\, \bA\bB^T = \Id \scalp \bA\bB^T,
\eeq
where the usual symbols denoting the identity, the transpose, and the trace operator have been employed. 
In addition to the usual tensorial product between (vectors and) second-order tensors $\bA$ and $\bB$ 
\beq
\left(\bA \otimes \bB\right)[\bC] = (\bB\scalp\bC)\,\bA,
\eeq
for every $\bC$, we will make use of the two tensorial products 
\beq
\left(\bA \bob \bB\right)[\bC] = 
\bA\frac{\bC + \bC^T}{2}\bB^T,~~~\left(\bA \gdp \bB\right)[\bC] = \bA\bC\bB^T,
\eeq
so that $\Id\bob\Id$ and $\Id \gdp \Id$ become the symmetrizing and the identity fourth-order tensors, respectively.

\section{A simple constitutive model evidencing flutter instability}

We refer here to the model proposed by Bigoni and Petryk (2002) as a large strain version of that proposed by
Bigoni and Loret (1999) [see also Bigoni (1995) and Bigoni and Zaccaria (1994)].
In particular, an objective symmetric flux, namely, the Oldroyd derivative of the Kirchhoff stress 
\beq
\lb{oldr}
\oldr{\bK} = \dot{\bK} - \bL \bK - \bK \bL^{T},
\eeq
(where a dot over a symbol denotes material time derivative, 
$\bL = \dot{\bF} \bF^{-1}$ is the spatial velocity gradient and $\bF$ the deformation gradient) 
is related to the Eulerian strain rate 
\beq
\bD = \frac{1}{2} \left(\bL + \bL^T \right) ,
\eeq
through the piecewise-linear elastoplastic constitutive equation
\beq
\lb{plastboy}
\oldr{\bK} = 
\left\{ 
\barr{ll}
\ds
\fE[\bD] - \frac{1}{H} \left< \bQ \scalp \fE[\bD] \right> \fE[\bP] & ~~~ \mathrm{if}~f(\bK,\mK) = 0, \\[5mm]
\fE[\bD] & ~~~ \mathrm{if}~f(\bK,\mK) < 0,
\earr
\right.
\eeq
where the symbol $\left< \scalp \right>$ denotes the Macaulay brackets operator (defined for every scalar 
$\alpha$ as $\left< \alpha \right> = (\alpha + |\alpha|) / 2$), $\fE$ is the elastic fourth-order tensor, $f$ is 
the yield function in stress space depending on a collection $\mK$ of internal variables (of arbitrary scalar or 
tensorial nature); moreover, $\bP$ and $\bQ$ are the normals to the plastic potential and yield surface, 
respectively, and the plastic modulus $H$ is related to the hardening modulus $h$ through
\beq
H = h + \bQ \scalp \fE[\bP].
\eeq

In the present article, we will refer to the loading branch of eqn. (\ref{plastboy}), which is
\beq
\lb{loadb}
\dot{\bK} = \fE[\bL] + \bL \bK + \bK \bL^{T} - \frac{1}{H} (\fE[\bP] \otimes \fE^T[\bQ])[\bL],
\eeq
where we have used the minor symmetries of $\fE$. Finally, introducing the first Piola-Kirchhoff
stress
\beq
\bS = \bK\bF^{-T} , 
\eeq
eqn. (\ref{loadb}) can be rewritten as
\beq
\lb{eqcost}
\dot{\bS} = \fC[\dot{\bF}],
\eeq
where 
\beqar
\lb{opertan}
\lefteqn{
\fC = (\Id \gdp \bF^{-1}) \fE (\Id \gdp \bF^{-T}) + \Id \gdp \bF^{-1} \bS} \\ 
& & ~~~~~~~~~~~~~~~~~~~~~~~~~~~
- \frac{1}{H}(\Id \gdp \bF^{-1})(\fE[\bP] \otimes \fE[\bQ])(\Id \gdp \bF^{-T}). \nonumber
\eeqar
Note that 
the tangent constitutive operator $\fC$, eqn.~(\ref{opertan}), possesses neither the minor nor 
the major symmetry, the latter except in the associative case, $\bQ = \bP$.

\subsection{Anisotropic elasticity}

Following Bigoni and Loret (1999) an anisotropic elastic law is assumed in the form 
\beq
\lb{aniso}
\fE = \lambda \bB \otimes \bB + 2 \mu \bB \bob \bB,
\eeq
where $\lambda$ and $\mu$ are material constants subject to the restrictions $\mu > 0$, 
$3 \lambda + 2 \mu >0$, and $\bB$ is a symmetric, positive definite second-order tensor, selected in the format
\beq
\bB = b_1 \bb \otimes \bb + b_2 (\Id - \bb \otimes \bb),
\eeq
where $b_1$ and $b_2$ are the eigenvalues of $\bB$, while 
the line spanned by the unit vector $\bb$ and the plane perpendicular to it are the corresponding eigenspaces.
Moreover, the material constants $b_1$ and $b_2$ are assumed to depend on a single angular parameter $\hat{b}$, 
restricted to the range $]0^{\circ},90^{\circ}[$ to meet the positive definiteness requirement of $\bB$, 
\beq
\lb{b12}
b_1 = \sqrt{3} \, \cos{\hat{b}},~~~ b_2 = \sqrt{\frac{3}{2}} \, \sin{\hat{b}},
\eeq
so that the isotropic behaviour is recovered when $b_1 = b_2 = 1$, or $\hat{b} \approx 54.74^{\circ}$.

\subsection{ The acoustic tensor}

The acoustic tensor $\bA^{ep}(\bn)$ associated with the tangent constitutive operator $\fC$ and the 
mass density $\rho$ is defined by 
\beq
\bA^{ep}(\bn) \bg = \frac{1}{\rho} \, \fC[\bg \otimes \bn] \bn,
\eeq
where $\bn$ and $\bg$ are the direction and amplitude of the propagating wave, respectively. 
Therefore, the acoustic tensor corresponding to $\fC$ in eqn.~(\ref{opertan}) is
\beq
\lb{tenac}
\bA^{ep}(\bn) = \bA^{e}(\bn) - \frac{1}{\rho \, H} \left( \fE[\bP] \bF^{-T} \bn \otimes \fE[\bQ] \bF^{-T} \bn \right),
\eeq
where $\bA^{e}(\bn)$ is the elastic acoustic tensor, defined as
\beqar
\lb{tenacel}
\lefteqn{
\bA^{e}(\bn) = \frac{\lambda + \mu}{\rho}(\bB \bF^{-T} \bn) \otimes (\bB \bF^{-T} \bn)} \\ 
& & ~~~~~~~~~~~~~~~
+ \frac{\mu}{\rho} \left[ (\bF^{-T} \bn) \scalp (\bB \bF^{-T} \bn) \right] \bB 
+ \frac{1}{\rho} \left[ \bn \scalp (\bF^{-1} \bS \bn) \right] \Id .\nonumber
\eeqar
Since $\fC$ does not have the major symmetry, the acoustic tensor (\ref{tenac})--(\ref{tenacel}) 
is also not symmetric.

\subsection{Examples of flutter instability for plane problems}
\lb{planar}

The current configuration is assumed as reference, so that $\bF = \Id$ and $\bS = \bK = \bT$, where $\bT$
denotes the Cauchy stress.  
The plane problem is considered in which vector $\bb$ and the propagation direction $\bn$ lie in the plane spanned 
by $\bk_1$ and $\bk_2$, two unit eigenvectors of $\bK = \bT$. Assuming the 
Drucker-Prager yield criterion, tensors $\bP$ and $\bQ$ take the form
\beq
\lb{druc}
\bP = \cos \chi \frac{\dev \, \bT}{|\dev \, \bT|} + \frac{\sin \chi}{\sqrt{3}}\, \Id,~~~
\bQ = \cos \psi \frac{\dev \, \bT}{|\dev \, \bT|} + \frac{\sin \psi}{\sqrt{3}}\, \Id,
\eeq
respectively, where $\dev \, \bT = \bT - \tr \bT/3$ and  
the angular parameters $\chi$ and $\psi$ describe respectively the dilatancy and the pressure-sensitivity 
of the material.

In the reference system $\{\bn,\bs,\bk_3\}$, where $\bs = \bk_3 \times \bn$, the acoustic tensor $\bA^{ep}(\bn)$
becomes
\beq
\lb{matrix}
\left(
\barr{ccc}
\ds A^{e}_{nn} - \frac{1}{\rho H}(\bn \scalp \bq)(\bn \scalp \bp) & 
\ds A^{e}_{ns} - \frac{1}{\rho H}(\bn \scalp \bq)(\bs \scalp \bp) & 0 \\[5mm]
\ds A^{e}_{ns} - \frac{1}{\rho H}(\bs \scalp \bq)(\bn \scalp \bp) & 
\ds A^{e}_{ss} - \frac{1}{\rho H}(\bs \scalp \bq)(\bs \scalp \bp) & 0 \\[5mm]
0 & 0 & \ds{\frac{\mu\,b_2 (\bn \scalp \bB\bn) + \bn \scalp \bT\bn}{\rho}}
\earr
\right),
\eeq
where 
\beq
\barr{l}
\bq \equiv \fE[\bQ] \bn = \lambda (\bB \scalp \bQ) \bB\bn + 2\mu \bB\bQ\bB\bn, \\[5mm]
\bp \equiv \fE[\bP] \bn = \lambda (\bB \scalp \bP) \bB\bn + 2\mu \bB\bP\bB\bn,
\earr
\eeq
and $A^{e}_{nn}$, $A^{e}_{ss}$, $A^{e}_{ns}$ are the in-plane components of the elastic acoustic tensor 
$\bA^{e}(\bn)$, namely
\beq
\barr{l}
A^{e}_{nn} = \ds{\frac{\lambda + 2 \mu}{\rho}(\bn \scalp \bB\bn)^2 + \frac{1}{\rho}
\bn \scalp \bT \bn}, \\[5mm]
A^{e}_{ss} = \ds{\frac{\lambda + \mu}{\rho}(\bs \scalp \bB \bn)^2 + \frac{\mu}{\rho}(\bn \scalp \bB \bn)(\bs \scalp \bB \bs) }
+ \ds{\frac{1}{\rho}\bn \scalp \bT \bn}, \\[5mm]
A^{e}_{ns} = \ds{\frac{\lambda + 2 \mu}{\rho}(\bn \scalp \bB \bn)(\bs \scalp \bB \bn)}.
\earr
\eeq
Note that the out-of-plane eigenvalue $A^{ep}_{33}$ in eqn. (\ref{matrix}) 
corresponds to a wave with 
out-of-plane amplitude ($\bg$ proportional to $\bk_3$) and is assumed to remain strictly positive.

From matrix (\ref{matrix}), we get the sum and the product of the two in-plane eigenvalues (squares of
the acceleration waves propagation velocities)
$c^{2}_1$ and $c^{2}_2$ 
corresponding to waves with in-plane amplitude ($\bg$ lying in the plane spanned by $\bk_1$ and $\bk_2$), 
\beq
\lb{sumprod}
\barr{l}
\ds c^{2}_1 + c^{2}_2 = A^{e}_{nn} + A^{e}_{ss} - \frac{1}{\rho H}(f_1 - f_2), \\[5mm]
\ds c^{2}_1 c^{2}_2 = A^{e}_{nn}A^{e}_{ss} - (A^{e}_{ns})^2 
+ \frac{1}{\rho H}(A^{e}_{ns} f_3 - A^{e}_{ss} f_1 + A^{e}_{nn} f_2),
\earr
\eeq
where
\beq
\barr{l}
f_1 = (\bn \scalp \bq)(\bn \scalp \bp), ~~~ f_2 = - (\bs \scalp \bq)(\bs \scalp \bp), \\[2mm]
f_3 = (\bn \scalp \bq)(\bs \scalp \bp) + (\bs \scalp \bq)(\bn \scalp \bp).
\earr
\eeq
A necessary and sufficient condition for the existence of complex conjugate eigenvalues $a^{ep}_1$ and 
$a^{ep}_2$ is represented by the simultaneous fulfillment of the following three conditions 
(Bigoni and Loret, 1999)
\beq
\lb{cond}
\barr{l}
f_4 = (A^{e}_{nn} - A^{e}_{ss})^2 \left[ (f_1 + f_2 + 2 e f_3)^2 - (1 + 4 e^2)(f_1 - f_2)^2 \right] > 0, \\[5mm]
f_5 = (A^{e}_{nn} - A^{e}_{ss}) (f_1 + f_2 + 2 e f_3) > 0, \\[5mm]
\ds \frac{f_5 - \sqrt{f_4}}{(A^{e}_{nn} - A^{e}_{ss})^2 + 4 (A^{e}_{ns})^2} < 
\rho^2 H < \frac{f_5 + \sqrt{f_4}}{(A^{e}_{nn} - A^{e}_{ss})^2 + 4(A^{e}_{ns})^2},
\earr
\eeq
where 
\beq
e = \frac{A^{e}_{ns}}{A^{e}_{nn} - A^{e}_{ss}}.
\eeq

\begin{figure}[!htb]
\begin{center}
\vspace*{3mm}
\includegraphics[width=4.5cm]{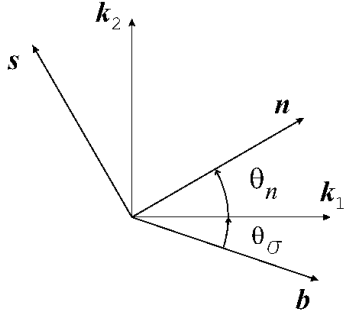}
\caption{\footnotesize Principal stress axes $\bk_1$ and $\bk_2$, axis of elastic symmetry $\bb$ and propagation 
direction $\bn$, singled out by angles $\theta_\sigma$ and $\theta_n$, respectively.}
\label{plane}
\end{center}
\end{figure}

With reference to Fig.~\ref{plane}, let $\theta_{\sigma}$ and $\theta_{n}$ be the angles of inclination of 
the direction of elastic anisotropy $\bb$ 
and wave propagation normal $\bn$ with respect to the stress principal axis $\bk_1$.

Dividing all quantities having the dimension of a stress in eqns.~(\ref{matrix})-(\ref{cond}) 
by $\mu$, the parameters on which the condition of flutter depends are:
\bit
\item Elastic parameters: $\lambda / \mu$, strength of anisotropy $\hat{b}$, 
and orientation of the axis of elastic symmetry with respect to the principal stress axis $\bk_1$,
namely, $\theta_{\sigma}$.
\item Plastic parameters: plastic modulus $H / \mu$, pressure sensitivity $\psi$, and dilatancy $\chi$ parameters.
\item Principal normalized deviatoric stress values: $\dev T_1 / |\dev \bT|$, $\dev T_2 / |\dev \bT|$, 
$\dev T_3 / |\dev \bT|$. However, these are not independent, 
so that given the form (\ref{druc}) of $\bP$ and $\bQ$, 
flutter depends on the angle 
\beq
\lb{lode}
\theta_L = \sgn{\left(\frac{\dev T_1}{|\dev \bT|}+2\frac{\dev T_2}{|\dev \bT|}\right)} 
\cos^{-1}\left(\sqrt{\frac{3}{2}} \frac{\dev T_1}{|\dev \bT|}\right) 
\eeq
in the deviatoric plane, which is a \lq modified Lode angle', defined for $\theta_L \in [-\pi, \pi]$ and in which 
$\sgn(0) = 1$.
\eit

\begin{figure}[!htb]
\begin{center}
\vspace*{3mm}
\includegraphics[width=6cm]{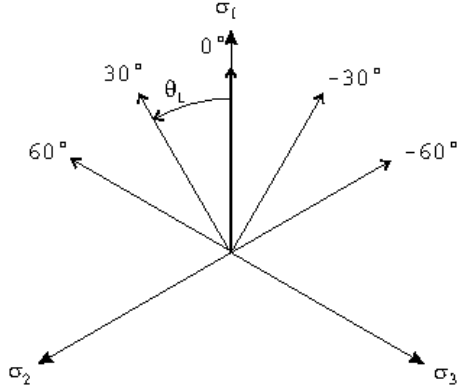}
\caption{\footnotesize Stress directions in the deviatoric plane, defined by the modified Lode angle (\ref{lode}), 
considered for flutter analysis.}
\label{stress}
\end{center}
\end{figure}

It is possible to study flutter for all the propagation 
directions $\bn$ while varying the plastic modulus $H / \mu$ and all 
remaining parameters in the above list are kept fixed, 
by use of inequalities (\ref{cond}). Therefore, 
the ranges in which flutter occurs can be plotted in the plane $H / \mu$ versus \ $\theta_{n}$. 
Restricting the analysis 
to the infinitesimal theory, where the flux (\ref{oldr}) is identified with $\dot{\bT}$, 
analyses have been performed for simplicity with different values of the modified Lode parameter $\theta_L$ 
$= \{60^\circ, 30^\circ, 0^\circ, -30^\circ, -60^\circ\}$, 
as indicated in Fig.~\ref{stress}.

\begin{figure}[!htb]
\begin{center}
\vspace*{3mm}
\includegraphics[width=16cm]{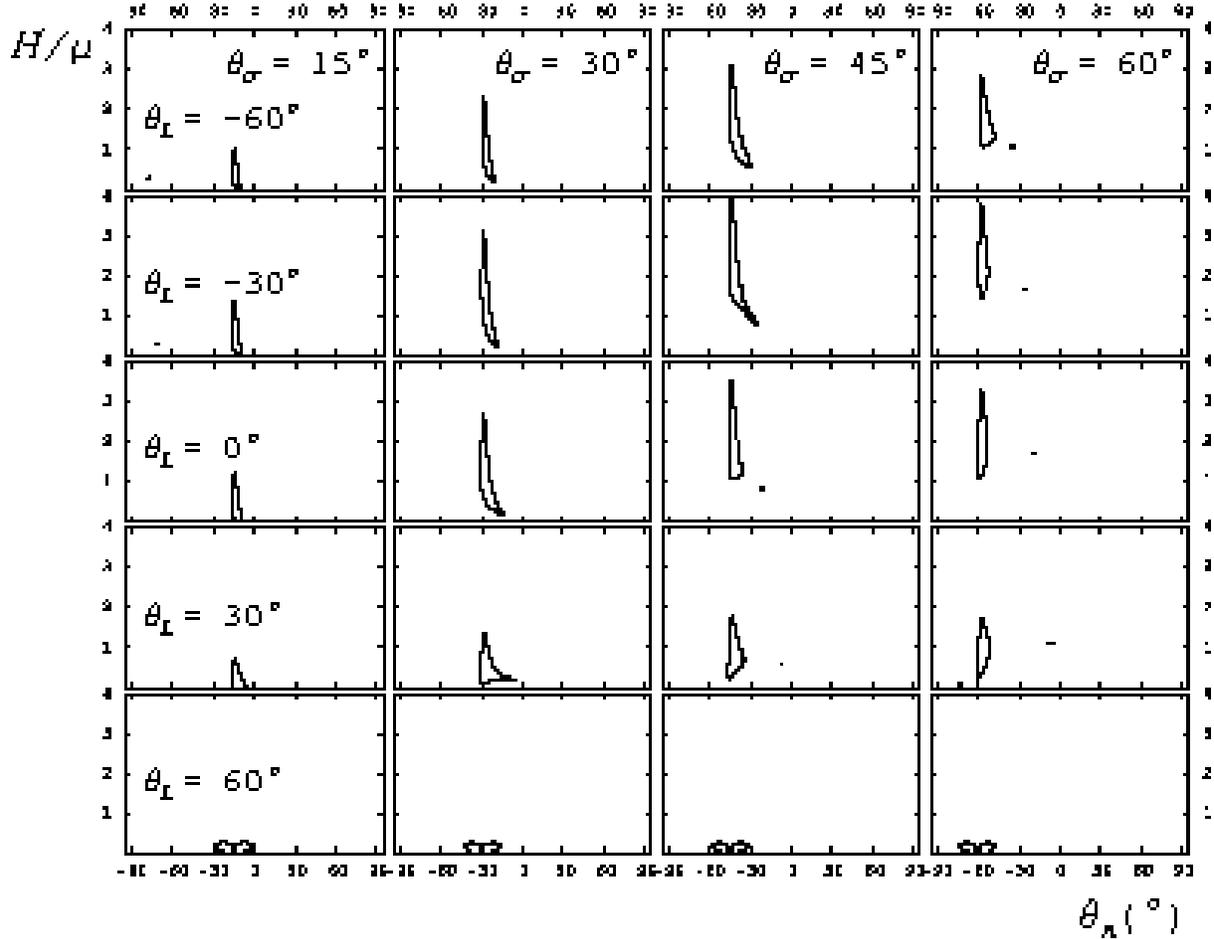}
\caption{\footnotesize Regions of flutter instability (occurring for internal points) 
in the $H / \mu$ vs.\ $\theta_{n}$ plane, 
for the stress paths shown in Fig. \ref{stress} at various anisotropy inclinations $\theta_{\sigma}$. 
The following values of material parameters have been considered: $\lambda / \mu = 1$, $\hat{b} = 80^{\circ}$, 
$\psi = 30^{\circ}$, and $\chi = 0^{\circ}$.}
\label{map}
\end{center}
\end{figure}

\begin{figure}[!htb]
\begin{center}
\vspace*{3mm}
\includegraphics[width=10cm]{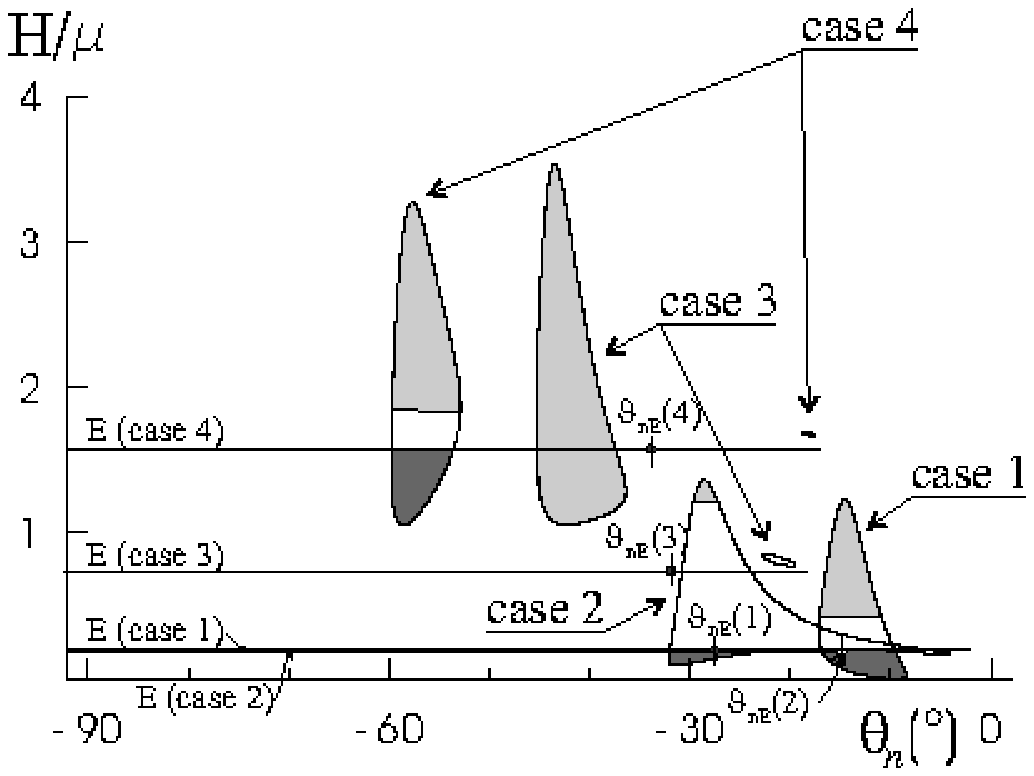}
\caption{\footnotesize Regions of flutter instability (occurring for internal points) 
in the $H / \mu$ vs.\ $\theta_{n}$ plane,
for $\lambda / \mu = 1$, $\hat{b} = 80^{\circ}$, $\psi = 30^{\circ}$, and $\chi = 0^{\circ}$.
Case 1: $\theta_L = 0^\circ$ and $\theta_{\sigma} = 15^{\circ}$.
Case 2: $\theta_L = 30^\circ$ and
$\theta_{\sigma} = 30^{\circ}$.
Case 3: as in case 1, but
$\theta_{\sigma} = 45^{\circ}$.
Case 4: as in case 1, but 
$\theta_{\sigma} = 60^{\circ}$. The regions of positive definiteness of the constitutive
operator are marked in light grey, while (E) denotes loss of ellipticity into shear bands (regions
shaded in dark grey) inclined at $\theta_{nE}(i)$, where i=1,..,4 denotes the relevant Case.}
\label{drops}
\end{center}
\end{figure}

Results are reported in Figs.~\ref{map} and \ref{drops}, the latter giving more detail for four of the cases 
reported in the former figure.
Different stress paths defined by the values of the modified Lode angle (\ref{lode}) reported in Fig. \ref{stress}
are considered for different anisotropy inclination $\theta_\sigma$ 
in  Fig.~\ref{map} at given values of $\psi = 30^\circ$ and $\chi = 0^\circ$. 
In the graphs the closed contours denote regions where flutter 
occurs in the plane defined by the normalized critical plastic modulus $H/\mu$ 
and the inclination of propagation  direction $\theta_n$. 

Four details of Fig.~\ref{map} are reported in 
Fig.~\ref{drops}, where 
$\lambda / \mu = 1$, $\hat{b} = 80^{\circ}$, $\psi = 30^{\circ}$, and $\chi = 0^{\circ}$, as in Fig.~\ref{map}.
The six regions in Fig.~\ref{drops} correspond to the four cases 
$\theta_L = 0^\circ$ and $\theta_{\sigma} = 15^{\circ}$ (Case 1), 
$\theta_L = \theta_{\sigma} = 30^{\circ}$ (Case 2),  
$\theta_L = 0$ and $\theta_{\sigma} = 45^{\circ}$ (Case 3), and 
$\theta_L = 0$ and $\theta_{\sigma} = 60^{\circ}$ (Case 4).

With reference to the Cases 1,2,3 and 4, detailed in Fig.~\ref{drops}, we note that 
the critical values of plastic modulus for loss of positive definiteness of the 
constitutive operator $H^{PD}_{cr}$ and for loss of ellipticity $H^{E}_{cr}$ permitting shear bands with normal inclined 
at $\theta_{n E}$ 
are\footnote{Note that with \lq ellipticity loss' we mean here the condition pertinent
to the underlying 
quasi-static deformation. Moreover, 
due to anisotropy, only one shear band is found as first noticed by Bigoni et al. (2000).}:
\beq
\lb{bottoni}
\begin{array}{lll}
\mbox{{\rm Case 1:}} & H^{PD}_{cr}/\mu = 0.42, & H^{E}_{cr}/\mu =  0.19, ~~~~\theta_{n E} = -28.0^\circ, \\ [5 mm]
\mbox{{\rm Case 2:}} & H^{PD}_{cr}/\mu = 1.22, & H^{E}_{cr}/\mu =  0.18, ~~~~\theta_{n E} = -16.4^\circ, \\ [5 mm]
\mbox{{\rm Case 3:}} & H^{PD}_{cr}/\mu = 1.03, & H^{E}_{cr}/\mu =  0.74, ~~~~\theta_{n E} = -32.0^\circ, \\ [5 mm]
\mbox{{\rm Case 4:}} & H^{PD}_{cr}/\mu = 1.84, & H^{E}_{cr}/\mu =  1.57, ~~~~\theta_{n E} = -33.9^\circ, \\ [5 mm]
\end{array}
\eeq
so that in all cases flutter may initiate when the constitutive operator is positive definite (therefore at an early 
stage of a deformation process) and may extend in a region possibly involving loss of ellipticity. 
Note that thresholds (\ref{bottoni}) have been graphically represented in Fig,~\ref{drops}, where 
light grey regions correspond to regions where flutter may occur with the constitutive operator still
positive definite, while in the dark grey regions ellipticity is lost (horizontal lines marking ellipticity loss are 
denoted with \lq E (case i)', where i = 1,..,4 stands for the number of the relevant Case). In the same figure, three black 
spots and a 
white spot (referred to Case 2) indicate the inclinations of shear bands at first loss of ellipticity.
Note that the small flutter regions of Cases 3 and 4 are beyond the positive definiteness threshold, 
but still in the elliptic region.
It may be important to remark that 
\begin{quote}
{\it the initial inclinations of propagation normals for flutter and shear bands are unrelated and 
remarkably different.}
\end{quote}

From the above analysis it can be deduced that the constitutive model allows one to approach flutter starting from a 
well-behaved state. Moreover, it may be interesting to note from Fig.~\ref{drops} that there are overlapping regions
corresponding to different stress states (Cases 1 and 2). In these zones the flutter may have identical characteristics
even if the stress state is different.

\subsection{Spectral analysis of the acoustic tensor}

The spectral analysis of the acoustic tensor is instrumental to the development
of the Green's function that will be presented in the next Section. 
The analysis is restricted to the in-plane components of the acoustic tensor $\bA^{ep}$ 
\beq
\lb{ttt}
\bA = A^{ep}_{11} (\bk_1 \otimes \bk_1) + A^{ep}_{12} (\bk_1 \otimes \bk_2) + A^{ep}_{21} (\bk_2 \otimes \bk_1) 
+ A^{ep}_{22} (\bk_2 \otimes \bk_2),
\eeq
represented for later convenience in the principal stress basis $\bk_1, \bk_2$.
The inverse of (\ref{ttt}) can be written as
\beqar
\lb{inversi}
\lefteqn{
\bA^{-1} = 
\frac{1}{A^{ep}_{11} A^{ep}_{22} - A^{ep}_{12} A^{ep}_{21}}
\left[ A^{ep}_{22} (\bk_1 \otimes \bk_1) - A^{ep}_{12} (\bk_1 \otimes \bk_2) \right.} \\
& & ~~~~~~~~~~~~~~~~~~~~~~~~~~~~~~~~~~~~
\left. - A^{ep}_{21} (\bk_2 \otimes \bk_1) + A^{ep}_{11} (\bk_2 \otimes \bk_2) \right]. \nonumber
\eeqar

Now, the eigenvalues of the acoustic tensor (\ref{ttt}) can be written in the form
\beq
\lb{eigenvalues}
\begin{array}{l}
c^2_1 \\ [5 mm]
c^2_2 
\end{array}
\left \} = 
\frac{A^{ep}_{11} + A^{ep}_{22} \pm \Delta}{2}, ~~~
\Delta = \sqrt{ (A^{ep}_{11} - A^{ep}_{22})^2 + 4 A^{ep}_{12} A^{ep}_{21} }, \right .
\eeq
so that assuming non-defectiveness, the spectral representations of $\bA$ and $\bA^{-1}$ are
\beq
\lb{pipi}
\bA = c^{2}_1 (\bv_1 \otimes \bw_1) + c^{2}_2 (\bv_2 \otimes \bw_2),
\eeq
and, assuming\footnote{For 
$\Delta \rightarrow 0$ (coalescence of the eigenvalues), the tensor $\bA$ becomes defective
(except for the 
trivial case where $\bA$ is isotropic) and each term in the spectral representation
of $\bA$, and also of 
$\bA^{-1}$, blows up but $\bA^{-1}$ continues to exist and to be defined correctly. Indeed a substitution of 
eqns.~(\ref{eigenvalues}) and (\ref{eigenvectors1}) or (\ref{eigenvectors2}) into 
eqn.~(\ref{inverse}) leads to eqn. (\ref{inversi}).
} $c^{2}_1 \neq 0$ and $c^{2}_2 \neq 0$, 
\beq
\lb{inverse}
\bA^{-1} = \frac{1}{c^{2}_1} (\bv_1 \otimes \bw_1) + \frac{1}{c^{2}_2} (\bv_2 \otimes \bw_2) ,
\eeq
where $\{\bv_1,\bv_2\}$ and $\{\bw_1,\bw_2\}$ are dual bases, thus satisfying $\bv_i \scalp \bw_j = \delta_{ij}$ 
($i,j = 1,2$), composed of right, $\bv_i$, and left, $\bw_i$, eigenvectors. This basis is given by
\beq
\lb{eigenvectors1}
\barr{ll}
\ds 
\bv_1 = \bk_1 + \frac{\Delta - (A^{ep}_{11} - A^{ep}_{22})}{2 A^{ep}_{12}} \bk_2, ~~~&
\ds 
\bv_2 = \bk_1 + \frac{-\Delta - (A^{ep}_{11} - A^{ep}_{22})}{2 A^{ep}_{12}} \bk_2, \\[5mm]
\ds 
\bw_1 = \frac{\Delta + (A^{ep}_{11} - A^{ep}_{22})}{2 \Delta} \bk_1 + \frac{A^{ep}_{12}}{\Delta} \bk_2, ~~~&
\ds 
\bw_2 = \frac{\Delta - (A^{ep}_{11} - A^{ep}_{22})}{2 \Delta} \bk_1 - \frac{A^{ep}_{12}}{\Delta} \bk_2,
\earr
\eeq
when $A^{ep}_{12} \neq 0$, or by
\beq
\lb{eigenvectors2}
\barr{ll}
\ds 
\bv_1 = \frac{\Delta + (A^{ep}_{11} - A^{ep}_{22})}{2 A^{ep}_{21}} \bk_1 + \bk_2, ~~~&
\ds 
\bv_2 = \frac{-\Delta + (A^{ep}_{11} - A^{ep}_{22})}{2 A^{ep}_{21}} \bk_1 + \bk_2, \\[5mm]
\ds 
\bw_1 = \frac{A^{ep}_{21}}{\Delta} \bk_1 + \frac{\Delta - (A^{ep}_{11} - A^{ep}_{22})}{2 \Delta} \bk_2, ~~~&
\ds 
\bw_2 = -\frac{A^{ep}_{21}}{\Delta} \bk_1 + \frac{\Delta + (A^{ep}_{11} - A^{ep}_{22})}{2 \Delta} \bk_2,
\earr
\eeq
when $A^{ep}_{21} \neq 0$. The case $A^{ep}_{21} = A^{ep}_{12} = 0$ is trivial.

\section{The dynamic time-harmonic Green's function for general nonsymmetric constitutive equations}

An initial static homogeneous deformation of an infinite body is considered, satisfying equilibrium in terms
of first Piola-Kirchhoff stress, namely, 
\beq
\div \bS = \b0,
\eeq
and taken as the reference state in an updated Lagrangian formulation.
A dynamic perturbation is superimposed upon this state, defined by an incremental displacement $\bu$ satisfying
the equations of incremental motion, written with reference to the constitutive equation (\ref{eqcost}) 
in which dotted symbols are to be interpreted now as incremental quantities rather than rates. Thus
\beq
\lb{quattro}
\fC_{ijkl} u_{k,lj} + f_i = \rho \, u_{i,tt},
\eeq
where $_{,t}$ denotes material time derivative and 
$f_i$ and $\rho$ are the incremental body forces and the mass density, respectively.

Equations (\ref{quattro}) look like ordinary elastodynamics, except that 
\begin{quote}
$\fC_{ijkl}$  {\it 
has neither the usual major } $\fC_{ijkl} \neq \fC_{klij}$ {\it nor the minor }
$\fC_{ijlk} \neq \fC_{ijkl} \neq \fC_{jikl}$ 
{\it symmetries}. 
\end{quote}
Note that tensor $\fC_{ijkl}$ can be identified (and will be in the examples) with that provided 
by eqn. (\ref{opertan}), but can also be thought completely arbitrary in the following.
To investigate the properties of 
eqn.~(\ref{quattro}), outside and inside the flutter region we follow the Bigoni and Capuani (2002; 2005)
approach, based on the determination of the dynamic Green's 
function, sought for simplicity under the time-harmonic assumption
\beq
u_i(\bx, t) = \hat{u}_i(\bx) e^{-i \omega t}, ~~~ f_i(\bx, t) = \hat{f}_i(\bx) e^{-i \omega t},
\eeq 
where $\omega$ is the circular frequency and $t$ and $\bx$ denote time and space variables, respectively, so 
that the time dependence can be removed from eqn.~(\ref{quattro}) and consequently
\beq
\lb{strn}
\fC_{ijkl} \hat{u}_{k,lj} + \rho \, \omega^2 \hat{u}_i + \hat{f}_i = 0.
\eeq
The Green's tensor $G_{ip}(\bx)$ is obtained by solving eqn. (\ref{strn}) under the
hypothesis $\hat{f}_i = \delta_{ip} \delta(\bx)$, with $\delta(\bx)$ denoting the Dirac delta.
We obtain
\beq
\lb{green}
\fC_{ijkl} G_{kq,lj}(\bx) + \rho \, \omega^2 G_{iq}(\bx) + \delta_{iq} \delta(\bx) = 0.
\eeq

In order to approach the flutter condition, we exploit the analysis of the acoustic tensor developed for the 
planar problem in Section \ref{planar}, considering an infinite medium subject to 
plane strain (or generalized plane stress conditions), in which only four relevant components of
the Green's function appear
\beq
G_{iq} = G_{iq}(x_1,x_2),~~~i,\,q = \{1,2\},
\eeq
and depend only on the two coordinates $x_1$ and $x_2$.

\subsection{Radon transform}
The Green's function is determined employing a Radon transform 
technique [the alternative approach employed by Bigoni and Capuani (2005) and based on a plane wave
expansion is presented for completeness in Appendix A]. 
The Radon transform of a generic function $f(\bx)$, $\bx \in \Reals^2$ is defined as 
\beq
\mR \left[ f(\bx) \right] = 
\hat{f}(p,\bn) = \int_{\Reals^2} f(\bx) \delta(p - \bn \scalp \bx)\, d\bx, ~~~ p \in \Reals, ~ \bn \in \Reals^2
\eeq
with the inverse
\beq
f(\bx) = 
\frac{1}{4 \pi^2} 
\int_{|\bn|=1} \pv_{-\infty}^{+\infty} \frac{\hat{f}'(p,\bn)}{(\bn \scalp \bx - p)}\, dp\, ds,
\eeq
where a prime denotes partial differentiation in the following way
\beq
\hat{f}'(p,\bn) = \deriv{\hat{f}(p,\bn)}{p}.
\eeq
In addition to the linearity, we will make use of the following properties of the Radon transform:
\bit
\item derivative transforms
\beq
\mR \left[ f_{,j}(\bx) \right] = n_j \hat{f}'(p,\bn), ~~~
\mR \left[ f_{,lj}(\bx) \right] = n_l n_j \hat{f}''(p,\bn),
\eeq
\item transform of the two-dimensional Dirac delta function
\beq
\mR \left[ \delta(\bx) \right] = \delta(p).
\eeq
\eit
The Radon transform of eqn.~(\ref{green}) is therefore
\beq
\lb{green3}
\fC_{ijkl} n_l n_j \hat{G}''_{kq}(p,\bn) + \rho \, \omega^2 \hat{G}_{iq}(p,\bn) + \delta_{iq} \delta(p) = 0,
\eeq
where
\beq
\hat{G}''_{kq}(p,\bn) = \nderiv{}{p}{2} \hat{G}_{kq}(p,\bn).
\eeq
Eqn.~(\ref{green3}) can be rewritten in tensorial form as
\beq
\lb{pep}
\bA(\bn) \hat{\bG}''(p,\bn) + \omega^2 \hat{\bG}(p,\bn) + \ds{\frac{\delta(p)}{\rho}} \Id = \b0.
\eeq

Let us assume that $\bA(\bn)$ has two non-null and distinct eigenvalues $c_N^2$ and corresponding left and right 
eigenvectors $\bw_N$, $\bv_N$, ($N = 1,2$), which can be used as dual basis vectors, therefore
satisfying $\bv_N \scalp \bw_M = \delta_{NM}$, ($N,M = 1,2$). Employing the spectral representations 
of $\bA(\bn)$ and $\Id$ 
\beq
\lb{rapspet}
\bA(\bn) = \sum_{N = 1}^{2} c_N^2 \bv_N \otimes \bw_N, ~~~~~ 
\Id = \sum_{N = 1}^{2} \bv_N \otimes \bw_N,
\eeq
in eqns.~(\ref{pep}) and representing the transformed Green's function as
\beq
\lb{oca}
\hat{\bG}(p,\bn) = \sum_{N = 1}^{2} \phi_N(p,\bn) \bv_N \otimes \bw_N,
\eeq
where $\phi_N$ is a (for the moment unknown) function of $p$ and $\bn$, we get
\beq
\lb{morta}
\sum_{N = 1}^{2} 
\left[ c_N^2 \phi_N'' + \omega^2 \phi_N + \ds{\frac{\delta(p)}{\rho}} \right] \bv_N \otimes \bw_N  = \b0,
\eeq
which is equivalent to the following uncoupled system of two equations,
\beq
\lb{ordinary3}
\phi_N'' + k_N^2 \phi_N + \frac{1}{\rho \, c_N^2} \delta(p) = 0, ~~~ N = 1,2,
\eeq
where the wavenumber $k_N = \omega/c_N$ has been introduced.
Since we have chosen the harmonic time dependence to be of the form $e^{-i \omega t}$, the outgoing wave solution 
of (\ref{ordinary3}) in the $p$ coordinate is: 
\beq
\phi_N(p,\bn) = - \frac{e^{i k_N |p|}}{2 \rho \, i k_N c_N^2}, 
\eeq
so that
\beq
\lb{hatG2}
\hat{\bG}(p,\bn) = - \sum_{N = 1}^{2} \frac{e^{i k_N |p|}}{2 \rho \, i k_N c_N^2} \bv_N \otimes \bw_N.
\eeq
and
\beq
\hat{\bG}'(p,\bn) = - \sum_{N = 1}^{2} \frac{\sgn (p) e^{i k_N |p|}}{2 \rho \, c_N^2} 
\bv_N \otimes \bw_N.
\eeq
The antitransform of equation (\ref{hatG2}) leads to
\beq
\lb{integralazzo}
\bG(\bx) = 
- \frac{1}{4 \pi^2} \sum_{N = 1}^{2} \int_{|\bn| = 1} \int_{-\infty}^{+\infty} 
\frac{\sgn (p) e^{i k_N |p|}}{2 \rho \, c_N^2 (\bn \scalp \bx -p)} \bv_N \otimes \bw_N \,dp\,ds.
\eeq
The integral in the variable $p$ can be evaluated in the way shown in Appendix B, so that, employing the cosine 
and sine integral functions
\beq
\Ci(z) = \int_{+ \infty}^{z} \frac{\cos\, t}{t} dt,~~~ |\arg z| < \pi ~~~ \mathrm{and} ~~~ 
\Si(z) = \int_{0}^{z} \frac{\sin\, t}{t} dt ,
\eeq
the Green's function can be finally written in the form
\beqar
\lb{sederino}
\lefteqn{
\bG (\bx) = - \frac{1}{8 \pi^2} \sum_{N = 1}^{2} \int_{|\bn|=1} 
\left[ 2 \cos(k_N \bn \scalp \bx) \Ci(k_N |\bn \scalp \bx|) \right.} \\
& & ~~~~~~~~~~
\left. + 2 \sin(k_N \bn \scalp \bx) \Si(k_N \bn \scalp \bx) - i \pi \cos(k_N \bn \scalp \bx) \right] 
\frac{\bv_N \otimes \bw_N}{\rho \, c_N^2} ds. \nonumber
\eeqar
We introduce polar coordinates so that the position vector $\bx$ has modulus 
$r = |\bx|$ and is inclined at angle $\theta$ to the $x_1$-axis. Taking the unit vector $\bn$ inclined at 
$\alpha + \theta$ with respect to the $x_1$-axis (so that $\alpha$ is the angle between $\bx$ and $\bn$) and 
noting that $\cos(\cdot)\,\Ci(\cdot)$ and $\sin(\cdot)\, \Si(\cdot)$
are even functions, we can re-write eqn. (\ref{sederino}) as
\beqar
\lb{pera}
\lefteqn{
\bG (\bx) = - \frac{1}{8 \pi^2} \sum_{N = 1}^{2} \int_{0}^{2 \pi} 
\left[ 2 \cos(r k_N |\cos\,\alpha|) \Ci(r k_N |\cos\,\alpha|) \right.} \\
& & ~~~~~
\left. + 2 \sin(r k_N |\cos\,\alpha|) \Si(r k_N |\cos\,\alpha|) - i \pi \cos(r k_N |\cos\,\alpha|) \right] 
\frac{\bv_N \otimes \bw_N}{\rho \, c_N^2} d\alpha, \nonumber
\eeqar
where $k_N$, $\bv_N$, $\bw_N$ and $c_N^2$ depend on $\alpha + \theta$. 

The acoustic tensor is a periodic function of $\alpha$ with period $\pi$ since
\beq
\rho \, A_{ik}(\bn) =  \fC_{i1k1} n_1^2 + (\fC_{i1k2} +\fC_{i2k1}) n_1 n_2 + \fC_{i2k2} n_2^2,
\eeq
where $n_1 = \cos(\alpha+\theta)$ and $n_2 = \sin(\alpha+\theta)$, and 
also $c_N$, $k_N$, $\bv_N$, and $\bw_N$ are periodic 
functions of $\alpha$ with the same period. It follows that the integrand in eqn. (\ref{pera}) is $\pi$--periodic.
Therefore, 
\begin{quote}
{\em the two-dimensional, time-harmonic Green's function corresponding to a generic, completely 
non-symmetric constitutive fourth-order tensor,
relating the increment of the first Piola-Kirchhoff stress to the deformation
gradient increment, eqn. (\ref{eqcost})}, can be written in the form 
\end{quote}
\beqar
\lb{orzo}
\lefteqn{
\bG (\bx) = - \frac{1}{4 \pi^2} \sum_{N = 1}^{2} \int_{0}^{\pi} 
\left[ 2 \cos(r k_N |\cos\,\alpha|) \Ci(r k_N |\cos\,\alpha|) \right.} \\
& & ~~~~~
\left. + 2 \sin(r k_N |\cos\,\alpha|) \Si(r k_N |\cos\,\alpha|) - i \pi \cos(r k_N |\cos\,\alpha|) \right] 
\frac{\bv_N \otimes \bw_N}{\rho \, c_N^2} d\alpha, \nonumber
\eeqar
where  $k_N = \omega/c_N$ and $c_N^2$ are the eigenvalues of the acoustic tensor $\bA$, eqn. (\ref{pipi}) with 
corresponding left and right eigenvectors $\bw_N$ and $\bv_N$, all quantities depending on $\bn$, which means
on $\alpha + \theta$.

It can be noted that the integrand in eqn. (\ref{orzo}) displays a logarithmic 
singularity at $r=0$ and $\alpha = \pi / 2$, since (Lebedev, 1965)
\beq
\Ci(z) = \gamma + \log\, z - \int_{0}^{z} \frac{1 - \cos\, t}{t}\, dt, ~~~|\arg z| < \pi ,
\eeq
where $\gamma$ is Euler's constant.

\section{A dynamical interpretation of flutter instability} 

The dynamical interpretation of flutter instability will be achieved 
following the approach introduced by Bigoni and Capuani (2002; 2005), so that 
the Green's function is employed to provide
a dynamical perturbation to be superimposed upon a given state of equilibrium of a homogeneously
deformed material. Several plots 
of Green's tensor components will be presented, so that a preliminary 
normalization of the Green's tensor
and a study of the involved non-dimensional parameters becomes instrumental. In particular, 
introducing an arbitrary characteristic length $a$ and consequently the dimensionless spatial variable
$\bar{\bx} = \bx / a$, 
making use of the property 
\beq
\delta(a \bar{\bx}) = \frac{1}{a^2} \delta(\bar{\bx}), ~~~ \bar{\bx} \in \Reals^2,
\eeq
eqn. (\ref{green}) can be rewritten as 
\beq
\bar{\fC}_{ijkl} \frac{\partial^2 \bar{G}_{kq}(\bar{\bx})}{\partial \bar{x}_j \partial \bar{x}_l} 
+ \bar{\omega}^2 \bar{G}_{iq}(\bar{\bx}) 
+ \delta_{iq} \delta(\bar{\bx}) = 0, ~~~ \bar{\bx} \in \Reals^2
\eeq
where
\beq
\bar{\fC}_{ijkl} = \frac{\fC_{ijkl}}{\mu}, ~~~ 
\bar{\omega} = a \sqrt{\frac{\rho}{\mu}} \omega.
\eeq
Thus, a dimensionless version of the Green's tensor (\ref{sederino}) reads
\beqar
\lefteqn{
\bar{\bG}(\bar{\bx}) = 
- \frac{1}{8 \pi^2} \sum_{N = 1}^{2} \int_{|\bn|=1} 
\left[ 2 \cos(\bar{k}_N \bn \scalp \bar{\bx}) \Ci(\bar{k}_N |\bn \scalp \bar{\bx}|) \right.} \\
& & ~~~~~~~~~
\left. + 2 \sin(\bar{k}_N \bn \scalp \bar{\bx}) \Si(\bar{k}_N \bn \scalp \bar{\bx}) 
- i \pi \cos(\bar{k}_N \bn \scalp \bar{\bx}) \right]
\frac{\bv_N \otimes \bw_N}{\bar{c}_N^2}\, ds, \nonumber
\eeqar
where
\beq
\bar{k}_N = a k_N = \frac{\bar{\omega}}{\bar{c}_N}, ~~~ \bar{c}_N = \sqrt{\frac{\rho}{\mu}} c_N,
\eeq
so that $\bar{c}_N^2$ are the eigenvalues of the dimensionless acoustic tensor $\bar{\bA} = \rho \, \bA / \mu$.

\subsection{Effects of flutter instability on Green's tensor}

\begin{figure}[!p]
\begin{center}
\vspace*{3mm}
\includegraphics[width=12cm]{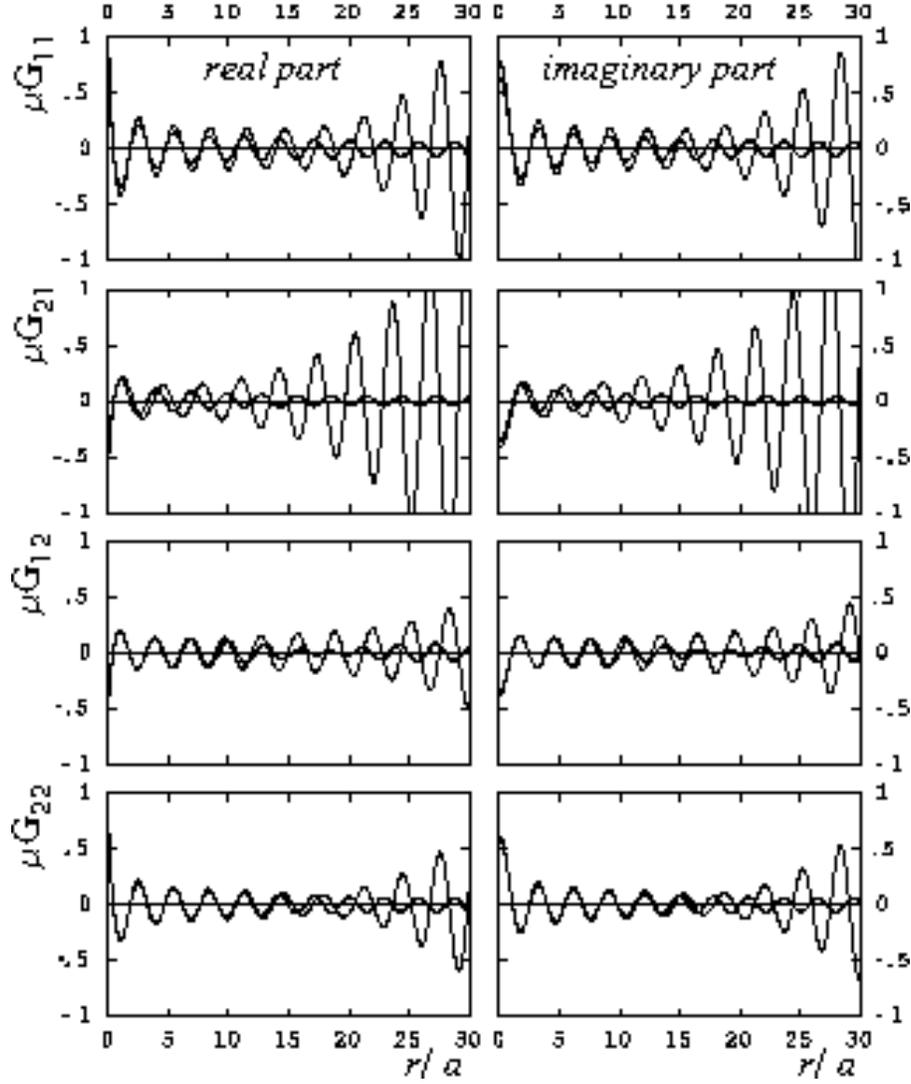}
\caption{\footnotesize Dimensionless Green's tensor components 
(real part left, imaginary part right in the figure) along a radial line inclined at $-45^\circ$ 
with respect to $x_1$-axis, 
for Case 3 of Fig.~\ref{drops} and 
$\bar{\omega} = 1$. 
Two values of the plastic modulus $H / \mu = \{3.53, 1.5\}$ 
are considered, corresponding, respectively, to situations near and inside the 
flutter region. The blow-up of all components of the Green's tensor 
is evident in the flutter region, $H / \mu = 1.5$. }
\label{greee}
\end{center}
\end{figure}

The behaviour of the Green's function, eqn. (\ref{orzo}), is briefly analyzed here,
outside and inside the 
flutter region.  
As a reference, we consider Case 3 shown 
in Fig.~\ref{drops}, in which the material is subject to the radial stress path corresponding
to $\theta_L = 0$ in Fig.~\ref{stress}
and the direction of the axis of elastic symmetry is taken inclined at 
$\theta_{\sigma} = 45^{\circ}$ with respect to the principal stress direction $\bk_1$. 
The employed material parameters 
are $\lambda / \mu = 1$, $\hat{b} = 80^{\circ}$, $\psi = 30^{\circ}$, and $\chi = 0^{\circ}$. 
The dimensionless Green's tensor 
components have been computed 
for $\bar{\omega} = 1$ and for several values of the plastic modulus $H / \mu$, including the 
values 3.53, and 1.5. These correspond, respectively, to 
situations near and inside the flutter region (see Fig.~\ref{drops}), but still in a situation
where the constitutive operator is positive definite.
The values of the components are plotted in Fig.~\ref{greee} as functions of the distance 
from the singularity along a radial line inclined at $-45^\circ$ with respect to the $x_1$ axis, 
normalized through division by $a$.
The real (imaginary) parts of the Green's function components are plotted left (right) in the 
figure, the plots having been obtained starting from $x_1 = 1/10$ to exclude the 
singularity (in the real components of the Green's tensor).

Commenting on the results, first, we note from the figure that the Green's tensor is not symmetric 
(since the acoustic tensor is not), so that $G_{12} \neq G_{21}$.

Second, results referring to values of plastic modulus $H/\mu$ higher than 3.53 and up to 7, not reported 
here for conciseness, produce curves practically coincident to those pertaining to $H/\mu = 3.53$; 
we can therefore conclude that there is not much difference between the situations in which the 
material is far from and very near to the flutter region. This feature has been
confirmed by us with several calculations (not reported here) and 
distinguishes flutter from shear banding, the latter becoming already visible when the condition of loss
of ellipticity is approached from the interior of the elliptic range (Bigoni and Capuani, 2002; 2005).

Third, a blow-up of the solution with the space variable, clearly visible in 
all components of the Green's tensor is the characteristic feature of instability inside the flutter region, 
$H/\mu = 1.5$. This 
blow-up is similar to that evidenced by Bigoni and Willis (1994), but in a constitutive setting including
viscosity, which is now absent.

It becomes evident that further exploration of flutter instability requires plotting of incremental 
displacement maps. These are obtained below employing a perturbation in the form of a pulsating dipole.

\subsection{Effects of flutter instability revealed by a perturbing dipole}

\begin{figure}[!htb]
\begin{center}
\vspace*{3mm}
\includegraphics[width=4cm]{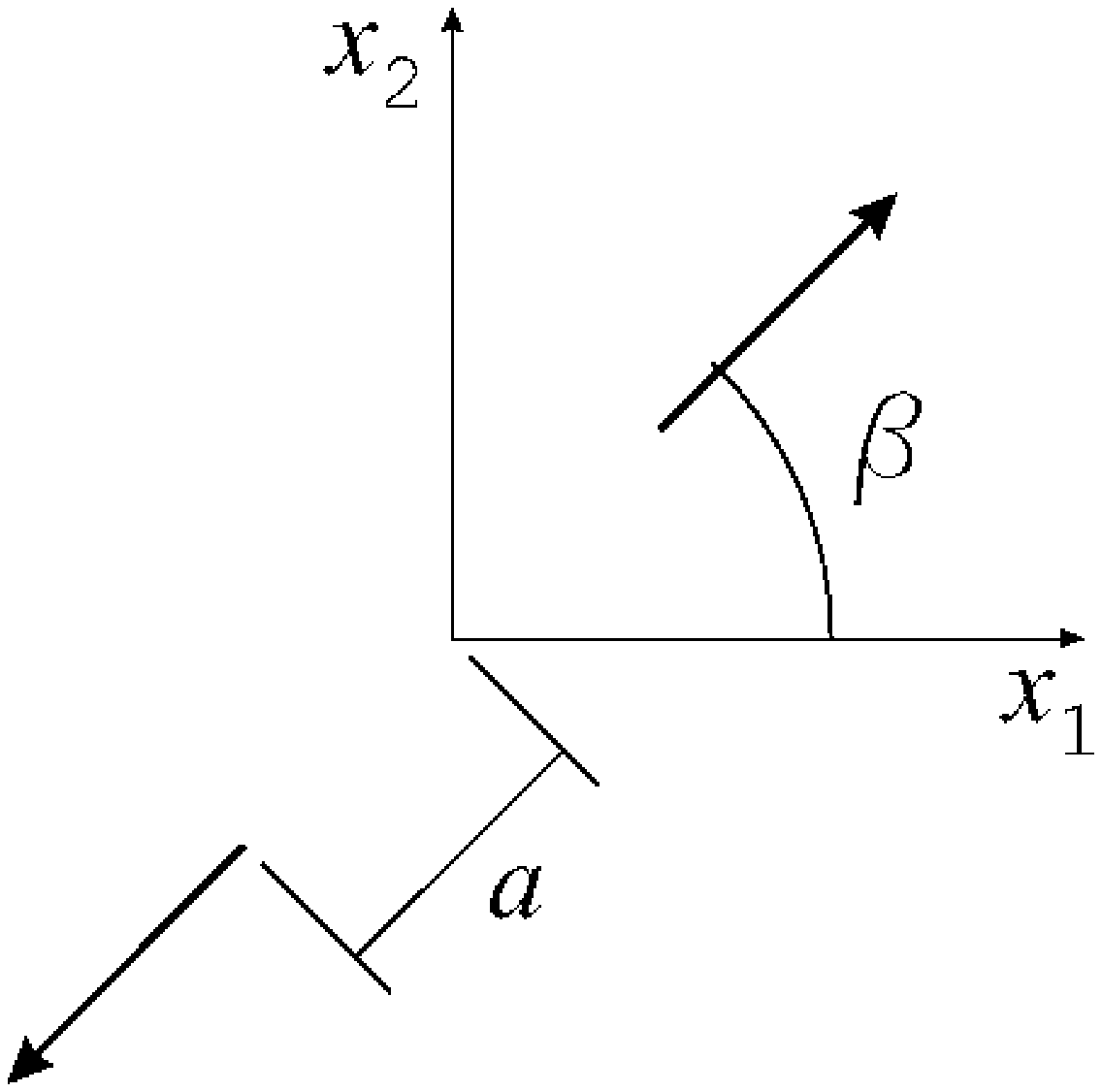}
\caption{\footnotesize Geometry of the time-harmonic pulsating perturbing dipole.}
\label{dipole}
\end{center}
\end{figure}

The singular solution previously obtained, eqn. (\ref{orzo}), 
can be used to analyze the effects of a perturbation superimposed upon 
a given homogeneous deformation of an infinite body. 
We follow here Bigoni and Capuani (2005) considering the simplest self-equilibrated perturbation 
in terms of a dipole: two equal and opposite pulsating forces of unit amplitude, taken 
at a distance $2 a$ apart, along a line inclined at $\beta = 45^{\circ}$ with respect to the $x_1$-axis, 
see Fig.~\ref{dipole}.

\begin{figure}[!p]
\begin{center}
\vspace*{3mm}
\includegraphics[width=10cm]{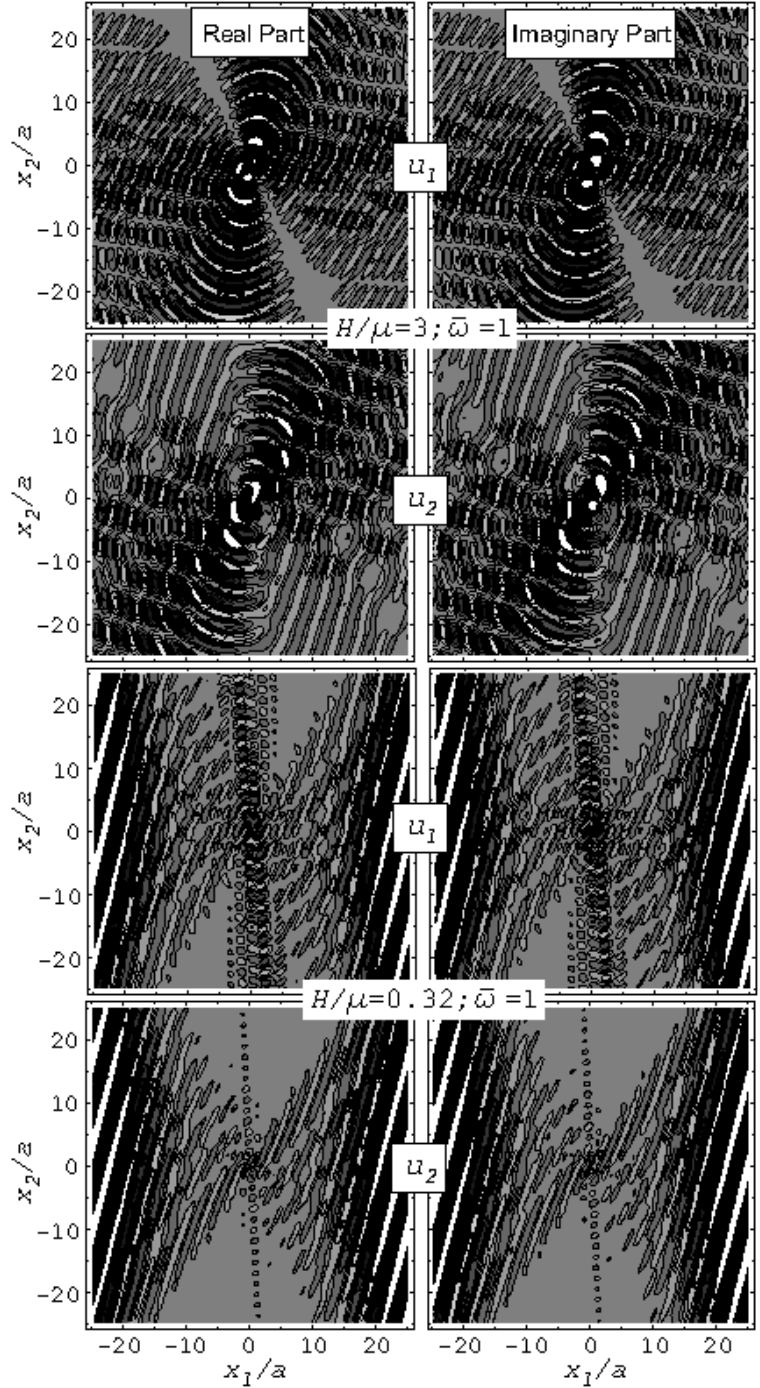}
\caption{\footnotesize Level sets of the real (left) and imaginary (right) parts of 
the components of incremental displacements ($u_1$ first and third parts from the top, $u_2$ second and fourth parts) for a dipole inclined at $\beta=45^{\circ}$, far from
(upper two parts, $H/\mu = 3$) and inside (lower two parts, $H/\mu = 0.32$) the
flutter region. Results pertain to Case 1 of Fig.~\ref{drops}, for $\bar{\omega} = 1$. Note the system of
blowing-up, parallel waves revealing the effect of flutter.
}
\label{mappe_caso_1}
\end{center}
\end{figure}

\begin{figure}[!p]
\begin{center}
\vspace*{3mm}
\includegraphics[width=10cm]{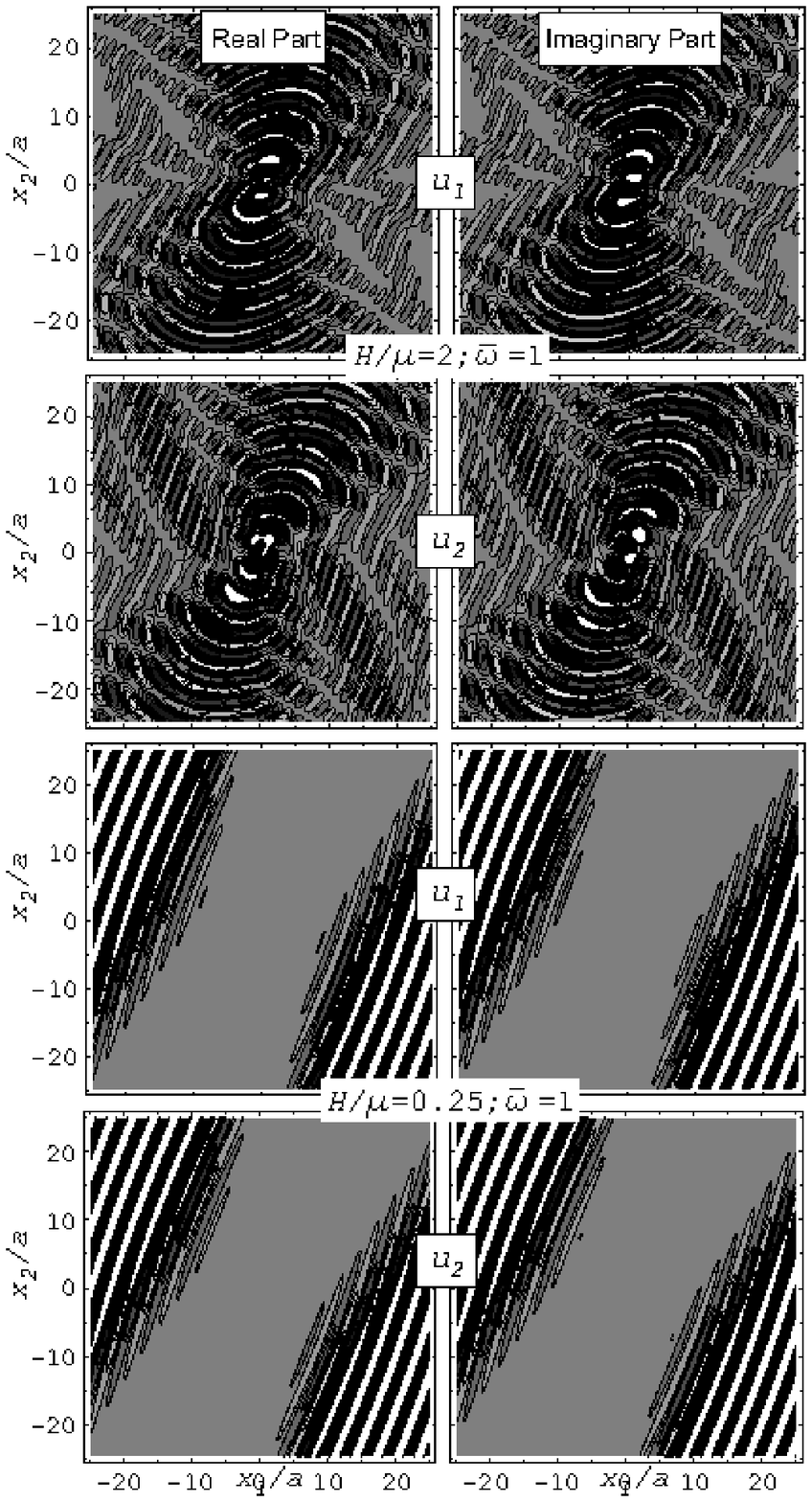}
\caption{\footnotesize Level sets of the real (left) and imaginary (right) parts of 
(the modulus of) incremental displacements for a dipole inclined at $\beta=45^{\circ}$, far from
(upper part, $H/\mu = 2$) and inside (lower part, $H/\mu = 0.25$) the
flutter region. Results pertain to Case 2 of Fig.~\ref{drops}, for $\bar{\omega} = 1$ .}
\label{mappe_caso_2}
\end{center}
\end{figure}

\begin{figure}[!p]
\begin{center}
\vspace*{3mm}
\includegraphics[width=10cm]{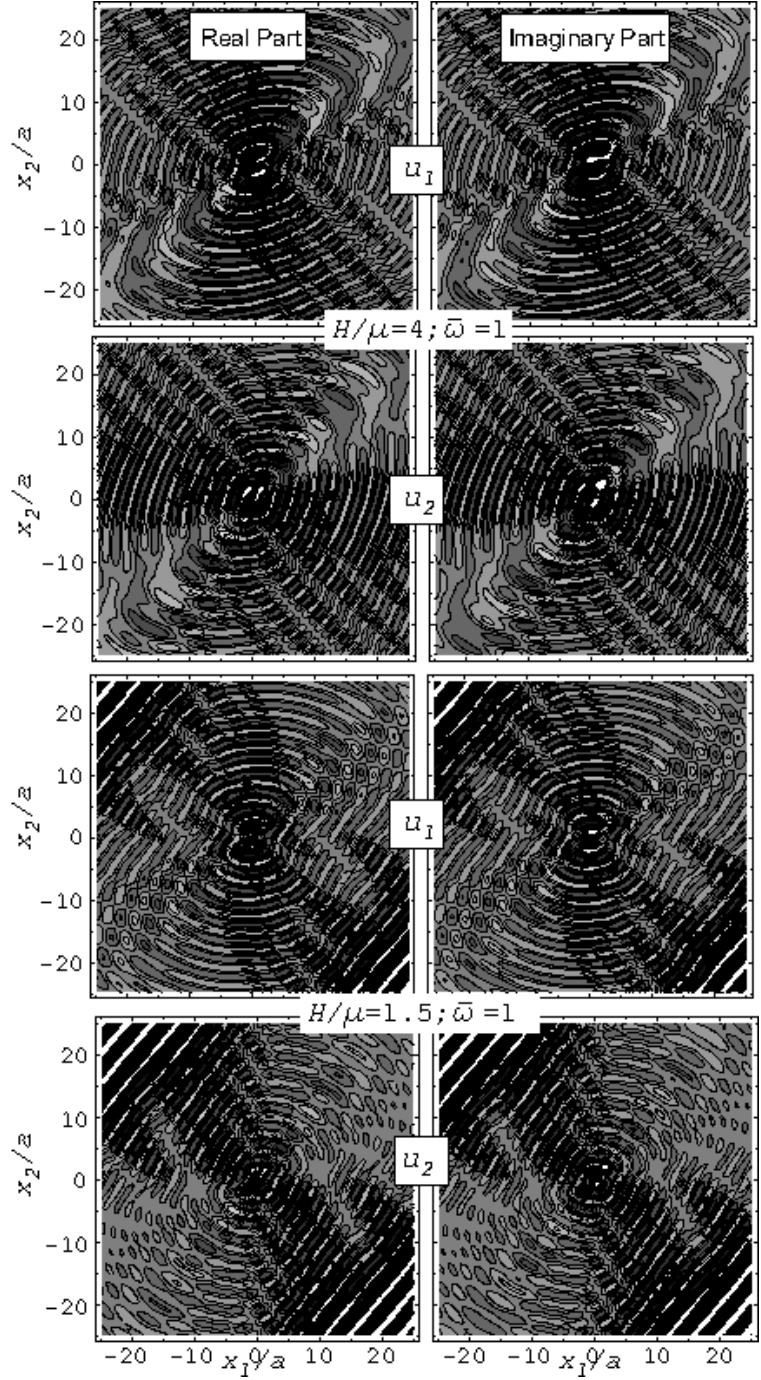}
\caption{\footnotesize Level sets of the real (left) and imaginary (right) parts of 
(the modulus of) incremental displacements for a dipole inclined at $\beta=45^{\circ}$, far from 
(upper part, $H/\mu = 4$) and inside (lower part, $H/\mu = 1.5$) the
flutter region. Results pertain to Case 3 of Fig.~\ref{drops}, for $\bar{\omega} = 1$ .}
\label{mappe_caso_3}
\end{center}
\end{figure}

\begin{figure}[!p]
\begin{center}
\vspace*{3mm}
\includegraphics[width=10cm]{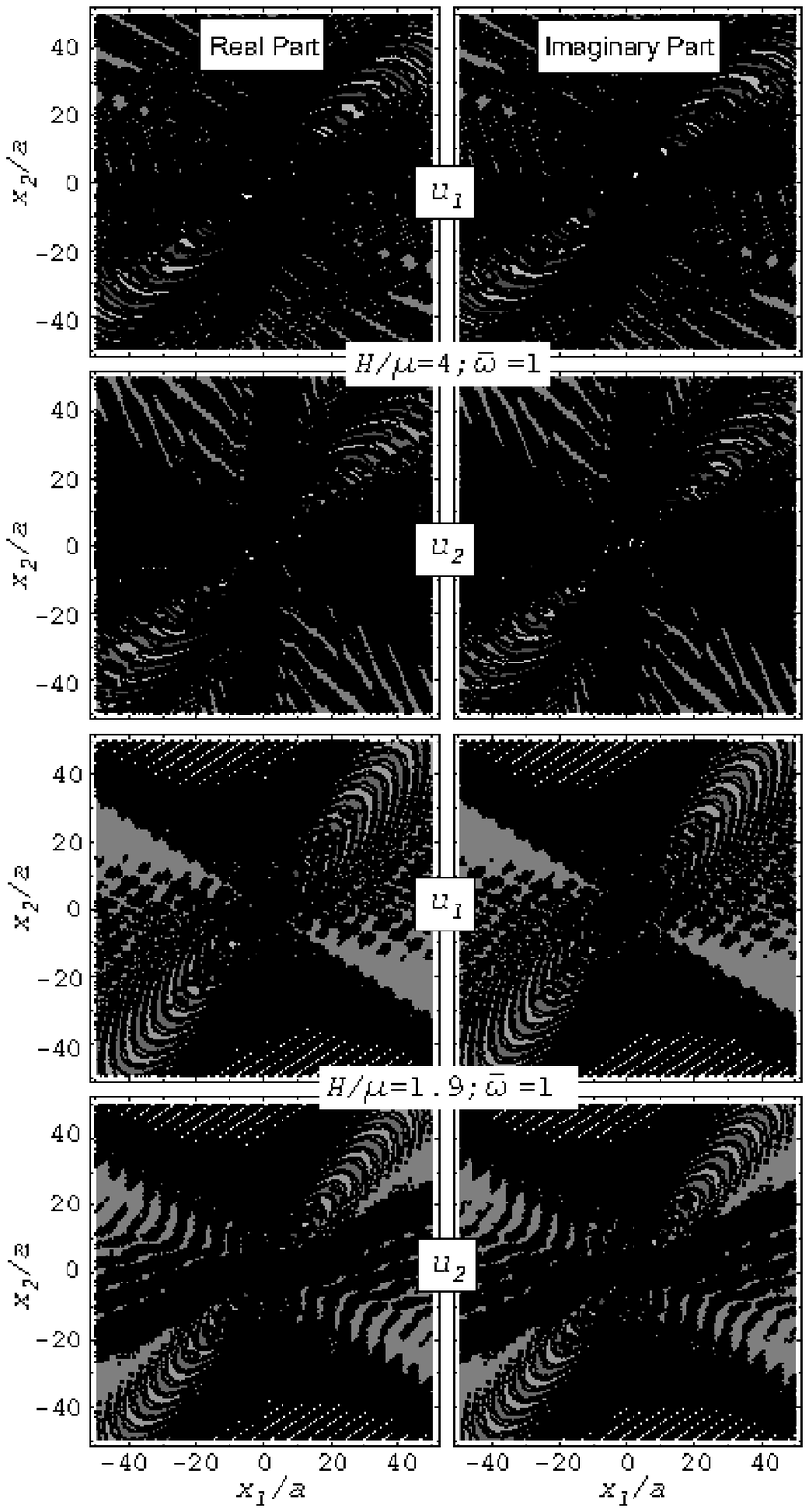}
\caption{\footnotesize Level sets of the real (left) and imaginary (right) parts of 
(the modulus of) incremental displacements for a dipole inclined at $\beta=45^{\circ}$, far from
(upper part, $H/\mu = 4$) and inside (lower part, $H/\mu = 1.9$) the
flutter region. Results pertain to Case 4 of Fig.~\ref{drops}, for $\bar{\omega} = 1$ .}
\label{mappe_caso_4}
\end{center}
\end{figure}

\begin{figure}[!p]
\begin{center}
\vspace*{3mm}
\includegraphics[width=10cm]{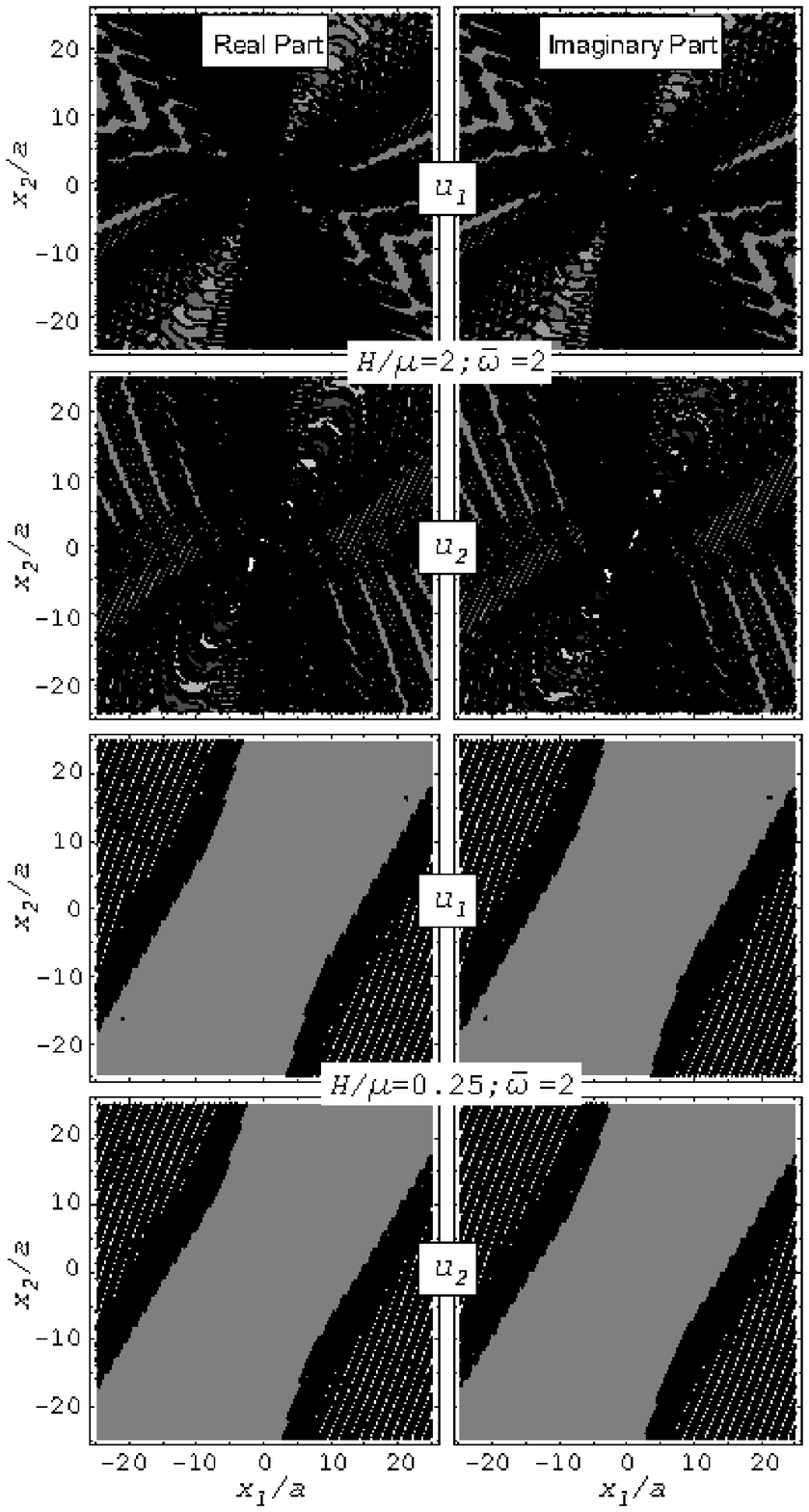}
\caption{\footnotesize As for Fig.~\ref{mappe_caso_2}, but with $\bar{\omega} = 2$ .}
\label{mappe_caso_2_freq_2}
\end{center}
\end{figure}

\begin{figure}[!p]
\begin{center}
\vspace*{3mm}
\includegraphics[width=10cm]{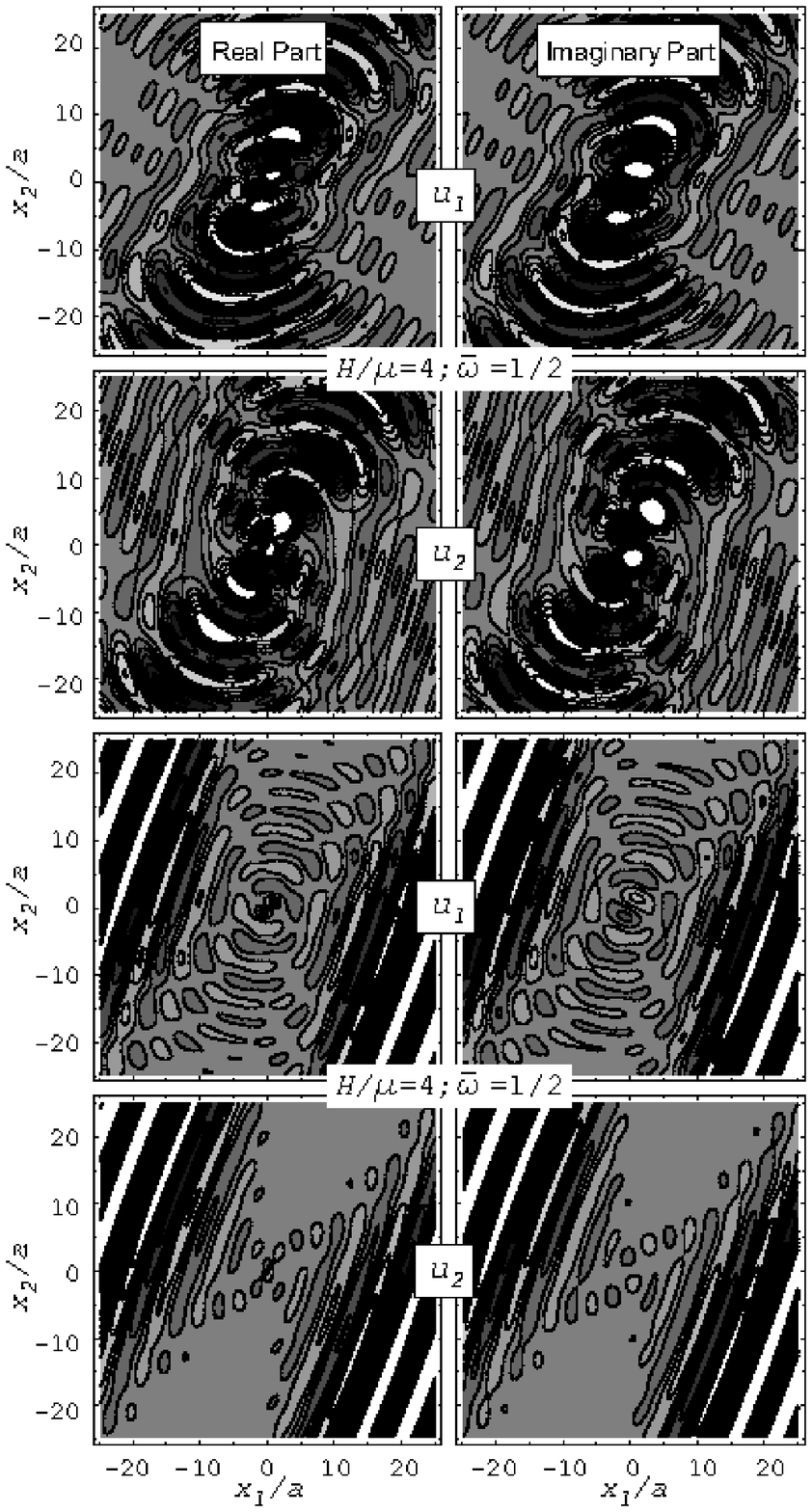}
\caption{\footnotesize As for Fig.~\ref{mappe_caso_2}, but with $\bar{\omega} = 1/2$ .}
\label{mappe_caso_2_freq_05}
\end{center}
\end{figure}

For this loading system, the level sets of the real part (left in the figures) and the 
imaginary part (right in the figures) of the components $u_1$ (first and third parts from the top of the figure) and $u_2$ 
(second and fourth parts from the top of the figure) 
of incremental displacements 
have been computed and plotted in Figs.~\ref{mappe_caso_1}--\ref{mappe_caso_2_freq_05}.
The two upper parts of all the figures refer to a situation far from flutter instability, 
whereas the two lower parts refer to a situation of flutter, well inside the region of instability.

The following parameters have been selected to be equal for all figures:
$$
\lambda/\mu = 1, ~~~ \hat{b}= 80^\circ, ~~~ \psi = 30^\circ, ~~~ \chi = 0^\circ.
$$
Moreover, Figs.~\ref{mappe_caso_1}--\ref{mappe_caso_4}
refer to the same nondimensional frequency parameter $\bar{\omega} = 1$, whereas
the effect of frequency is explored in Figs.~\ref{mappe_caso_2_freq_2} and  
\ref{mappe_caso_2_freq_05}, pertaining respectively to 
$\bar{\omega} = 2$ and $1/2$ and corresponding to the same parameters employed in 
Fig.~\ref{mappe_caso_2}. All components of incremental displacements have been plotted
for the nondimensional coordinates
$x_1/a$ and $x_2/a$ ranging between $-25$ and 25, with the exception of
Fig.~\ref{mappe_caso_4}, where this range 
has been extended to $-50$ and 50 to help visualization of the blowing-up typical of flutter.

The differences between Figs.~\ref{mappe_caso_1}--\ref{mappe_caso_4} lie in the choice 
of different stress states expressed in terms of $\theta_L$ and anisotropy direction 
$\theta_\sigma$. In particular: 
\begin{itemize}
\item Fig.~\ref{mappe_caso_1} refers to $H/\mu = 3$ (two upper parts), $H/\mu = 0.32$ (two lower parts) and to
Case 1 of Fig.~\ref{drops}, where $\theta_L=0^\circ$ and $\theta_\sigma= 15^\circ$;
\item Fig.~\ref{mappe_caso_2} refers to $H/\mu = 2$ (two upper parts), $H/\mu = 0.25$ (two lower parts) and to
Case 2 of Fig.~\ref{drops}, where $\theta_L=30^\circ$ and $\theta_\sigma= 30^\circ$;
\item Fig.~\ref{mappe_caso_3} refers to $H/\mu = 4$ (two upper parts), $H/\mu = 1.5$ (two lower parts) and to
Case 3 of Fig.~\ref{drops}, where $\theta_L=0^\circ$ and $\theta_\sigma= 45^\circ$;
\item Fig.~\ref{mappe_caso_4} refers to $H/\mu = 4$ (two upper parts), $H/\mu = 1.9$ (two lower parts) and to
Case 4 of Fig.~\ref{drops}, where $\theta_L=0^\circ$ and $\theta_\sigma= 60^\circ$.
\end{itemize}

Note that the values of the plastic modulus selected for the examples are all higher than the critical 
values for loss of ellipticity\footnote{More precisely, all the considered plastic moduli
are higher than the 
critical values for loss of {\it strong ellipticity} (Bigoni, 2000).} [see the values listed in (\ref{bottoni})], so 
that shear bands are excluded.
However, all the values of $H/\mu$ corresponding to situations far from flutter and the 
two values 1.5 and 1.9 lie in the zone of positive definiteness of the constitutive 
operator, while the two values 0.25 and 0.32 have been selected outside this region 
[see the values listed in (\ref{bottoni})].

It can be observed from the upper parts of Figs.~\ref{mappe_caso_1}--\ref{mappe_caso_4}
(referring to a non-flutter
situation) that the displacement maps are typical of an anisotropic material, since $45^\circ$--symmetry
is not in evidence. Moreover, decay of the solution is appreciable, when the distance from the 
dipole increases. 
Now, considering the lower parts of the figures, the effects of flutter instability become self-evident.
In particular, we may observe a growth of the solution in space, which tends to degenerate into a 
system of blowing-up, parallel plane waves.
Results not reported here for brevity demonstrate that: 
\begin{quote}
{\it the inclination of the blowing-up plane waves is almost independent of the dipole inclination (angle 
$\beta$ in Fig.~\ref{dipole}), so that it has to be considered a characteristic of the material, related
to the particular stress state and constitutive features. We have observed that the inclination of 
the blowing-up waves corresponds to
a value in the middle of the inclination fan of flutter (see Fig.~\ref{drops})}.
\end{quote}

In particular, the inclinations of the plane waves at a sufficient distance from the dipole, 
are different in Figs.~\ref{mappe_caso_1}-\ref{mappe_caso_4}, but correspond to the mean value of flutter direction fan
visible in Fig.~\ref{drops} at the analyzed $H/\mu$ values. On the other hand, the same inclinations are found 
for figures Figs.~\ref{mappe_caso_2} and \ref{mappe_caso_2_freq_2} and \ref{mappe_caso_2_freq_05}, since 
these cases differ only in the nondimensional frequency parameter $\bar{\omega}$, which influences only the 
spacing of the blowing-up waves.

As far as the 
effects of varying the nondimensional frequency parameter $\bar{\omega}$ are concerned (see 
Figs.~\ref{mappe_caso_2_freq_2} and \ref{mappe_caso_2_freq_05}, referring to the same material parameters
as in Fig.~\ref{mappe_caso_2}, but with $\bar{\omega} = \{1, 2, 1/2\}$),
we see that an increase in the frequency yields
a narrowing of the distance between blowing-up plane waves. Moreover, increase in frequency gives rise to
the \lq shadowing' effect already noted by Bigoni and Capuani (2005) for shear bands.

Compared to the shear bands analyzed by Bigoni and Capuani (2002; 2005), we may observe that these are
already revealed when the boundary of the region of ellipticity is approached from the inside, while 
flutter remains undetected. Beside this difference, there are however many similarities between the two phenomena:
first of all, shear bands tend to blow-up in space as the boundary of instability is approached, and 
extend from a perturbation to infinity, outside the elliptic 
range. 
Second, shear bands also tend to degenerate into families of plane waves parallel to a 
specific direction.
Third, the signals tend to focus along well defined patterns, both for shear bands and for flutter.
Note however, that flutter instability may 
occur much earlier than shear banding in a deformation process; moreover, waves near the loss of ellipticity threshold
tend to blow-up along the shear bands but, in contrast to flutter, they tend to decay 
in the parallel direction.

As a conclusion, we remark that flutter instability yields a self-organization of dynamic disturbances along
well-defined and blowing-up parallel waves, having inclinations corresponding to the mean value of the 
inclinations for which flutter is possible at the considered constitutive setting and stress state.

From the mechanical point of view, our results suggest that flutter yields 
a \lq layering' of deformation patterns, 
with an inclination corresponding to the flutter direction, 
a spacing related to the frequency of the perturbing agency, and possibly occurring 
early in a plastic deformation process.

\section{Conclusions}

Following the approach to material instabilities proposed by Bigoni and Capuani (2002; 2005), 
flutter instability
in a continuous elastoplastic medium has been investigated, by finding the dynamic, time-harmonic
Green's function for the loading branch of a fully unsymmetric tangent constitutive operator,
embodying features typical of the behaviour of granular materials. For this material, flutter
instability may occur when the constitutive operator is positive definite (so that the solution of the 
rate infinitesimal problem is unique and shear bands are excluded), while two eigenvalues
of the acoustic tensor are complex conjugate.
Our results provide the first interpretation of flutter instability, which is shown to 
correspond to a dynamical instability growing in space and self-organizing 
into plane waves with normals lying in the fan corresponding to the complex eigenvalues of 
the acoustic tensor and yielding a sort of \lq layering' of unstable deformation patterns, showing
some similarity to shear band instability. 
The rate of growth of the solutions displayed here increases with the frequency that is assumed.
This demonstrates dynamical ill-posedness of the governing equations of motion in the general
transient case and implies a need that is physical as well as mathematical for the admission
of some appropriate rate-dependence into the constitutive model, to remove the flutter effect
at high frequencies.  
Although no such mechanism is built into the present analysis (the tangent moduli would become functions of $\omega$
but this is in any case fixed), and other mechanisms not accounted for (such as for instance the possibility
of elastic unloading and material viscosity) may change some of our conclusions, we believe that the 
emergence of the layered structures that we have found may find future experimental validation.

\section*{Acknowledgments}

\noindent Financial support of MURST-Cofin 2004 (Microstructural problems and models:
applications in structural and civil engineering) is gratefully acknowledged.

\newpage
\setcounter{equation}{0}
\renewcommand{\theequation}{{A}.\arabic{equation}}
\begin{center}
{\bf APPENDIX A. Green's function obtained via plane wave expansion.}\\
\end{center}

The Green's function (\ref{sederino}) is obtained here for completeness using the plane wave expansion
technique employed by Bigoni and Capuani (2005).
The plane wave expansion of the $\delta$ function and of the Green's tensor $\bG(\bx)$ are, respectively,
\beq
\lb{peppa}
\delta(\bx) = - \frac{1}{4 \pi^2} \int_{|\bn|=1} \frac{1}{(\bn \scalp \bx)^2} ds,
~~~~
\bG(\bx) = - \frac{1}{4 \pi^2} \int_{|\bn|=1} \tilde{\bG}(\bn \scalp \bx) ds,
\eeq
where $\bn$ is a unit vector, so that the plane wave expansion of eqn.~(\ref{green}) leads to 
\beq
\fC_{ijkl} n_j n_l \tilde{G}_{kq}''(\xi) + \rho \, \omega^2 \tilde{G}_{iq}(\xi) + \frac{\delta_{iq}}{\xi^2} = 0,
\eeq
where $\xi = \bn \scalp \bx$. In this equation the acoustic tensor can be easily recognized, 
$A_{ik} = \fC_{ijkl} n_j n_l$, so that we get  
\beq
\lb{green2}
\bA(\bn) \tilde{\bG}''(\xi) + \omega^2 \tilde{\bG}(\xi) + \ds{\frac{1}{\rho \, \xi^2}} \Id = \b0.
\eeq 
Writing now the analogue of the representation (\ref{oca}), namely,
\beq
\lb{tildeg}
\tilde{\bG}(\xi) = \sum_{N = 1}^{2} \phi_N(\xi) \bv_N \otimes \bw_N,
\eeq
we transform eqn.~(\ref{green2}) into the analogue of eqn. (\ref{morta})
\beq
\sum_{N = 1}^{2} 
\left( c_N^2 \phi_N'' + \omega^2 \phi_N + \frac{1}{ \rho \,\xi^2} \right) \bv_N \otimes \bw_N = \b0,
\eeq
which is equivalent to the following uncoupled system of two equations, analogous to eqns. (\ref{ordinary3}),
\beq
\lb{ordinary2}
\phi_N'' + k_N^2 \phi_N + \frac{1}{\rho \, c_N^2} \frac{1}{\xi^2} = 0, ~~~ N=1,2,
\eeq
where $k_N = \omega / c_N$.

The sole physically meaningful solution of the ordinary differential equation (\ref{ordinary2}) is obtained by 
imposing the radiation condition, stating that the solution should include only outgoing waves. 
Since the harmonic time dependence has been selected in the 
form $e^{-i \omega t}$, the outgoing wave solution of (\ref{ordinary2}) in the $\xi$ coordinate is: 
\beqar
\lb{sole}
\lefteqn{
\phi_N(\xi) = 
\frac{1}{2 \rho \, c_N^2} \left[ 2 \Ci(k_N |\xi|) \cos(k_N \xi) \right.} \\
& & ~~~~~~~~~~~~~~~~~~~~~~~~~~~
\left. + 2 \Si(k_N \xi) \sin(k_N \xi) - i \pi \cos(k_N \xi) \right]. \nonumber
\eeqar

Finally, a chain of substitutions, of eqn.~(\ref{sole}) into 
eqn.~(\ref{tildeg}) and finally into eqn. (\ref{peppa})$_2$, leads to 
the Green's function in the form (\ref{sederino}).

\noindent


\vspace{10mm}

\setcounter{equation}{0}
\renewcommand{\theequation}{{B}.\arabic{equation}}
\begin{center}
{\bf APPENDIX B. Evaluation of the integral in the variable $p$ in eqn. (\ref{integralazzo}).}\\
\end{center}

The integral in the variable $p$ appearing in eqn. (\ref{integralazzo}) can be evaluated 
splitting the domain as follows
\beq
\int_{-\infty}^{+\infty} \frac{\sgn (p) e^{i k_N |p|}}{\xi - p}\, dp = 
- \int_{-\infty}^{0} \frac{e^{-i k_N p}}{\xi - p}\, dp
+ \int_{0}^{+\infty} \frac{e^{i k_N p}}{\xi - p}\, dp,
\eeq
so that we can treat the two integrals separately, namely
\beq
\lb{inteuno}
- \int_{-\infty}^{0} \frac{e^{-i k_N p}}{\xi - p}\, dp = 
- e^{-i k_N \xi} \int_{k_N \xi}^{+\infty} \frac{e^{i q}}{q}\, dq,
\eeq
where we have made the substitution $q = k_N (\xi - p)$, and
\beq
\lb{intedue}
\int_{0}^{+\infty} \frac{e^{i k_N p}}{\xi - p}\, dp = 
- e^{i k_N \xi} \int_{-k_N \xi}^{+\infty} \frac{e^{i q}}{q}\, dq,
\eeq
where we have made the substitution $q = k_N (p - \xi)$. The two expressions (\ref{inteuno}) and (\ref{intedue}) are 
used to get eqn. (\ref{sederino}).


\end{document}